\documentclass[sn-mathphys-num]{sn-jnl}
\usepackage{a4wide}
\usepackage[english]{babel}
\usepackage[normalem]{ ulem }
\usepackage{soul}
\usepackage{graphicx}
\usepackage{lipsum}

\usepackage[T1]{fontenc}
\usepackage[utf8]{inputenc}

\usepackage{amsmath,amsfonts,amssymb,amsopn,amscd,amsthm, mathrsfs}

\usepackage{gensymb}
\usepackage{comment}
\usepackage{dsfont}
\usepackage{bm}
\usepackage{graphicx}
\usepackage{xcolor}
\usepackage{epigraph}
\usepackage{enumerate}
\usepackage{xargs}
\usepackage[prependcaption]{todonotes}

\usepackage{marginnote}
\usepackage{todonotes}

\usepackage{hyperref}

\usepackage[charter]{mathdesign}

\usepackage{graphicx}
\usepackage[]{epsfig}
\usepackage[]{float}

\usepackage{tikz}
\usepackage{xcolor}
\usepackage{pstricks,pst-node}
\usetikzlibrary{shapes.misc,arrows,decorations.pathmorphing,backgrounds,positioning,fit,petri,shapes,petri,cd,arrows.meta,patterns,fadings,calc}

\newtheorem{theorem}{Theorem}[section]

\newtheorem{remark}[theorem]{Remark}

\numberwithin{equation}{section}
\begin{document}
 
 \title[MFT for superdiffusive systems: Ginzburg-Landau with long range interactions]{Macroscopic Fluctuation Theory for Ginzburg-Landau dynamics with long range interactions}
 
\author[1]{\fnm{C\'edric} \sur{Bernardin}}\email{sedric.bernardin@gmail.com}

\author*[2]{\fnm{Rapha\"el} \sur{Chetrite}}\email{Raphael.CHETRITE@univ-cotedazur.fr}

\affil[1]{\orgdiv{Faculty of Mathematics}, \orgname{National Research University Higher School of Economics},  \orgaddress{\street{6 Usacheva}, \city{Moscow}, \postcode{119048 }, \country{Russia}}}

\affil[2]{\orgdiv{Institut de Physique de Nice (INPHYNI)}, \orgname{Universit\'e C\^ote d'Azur, CNRS}, \orgaddress{\street{17 rue Julien Laupr\^etre}, \city{Nice}, \postcode{06200}, \country{France}}}

\abstract{Focusing on a famous class of interacting diffusion processes called Ginzburg-Landau (GL) dynamics, we extend the Macroscopic Fluctuations Theory (MFT) to these systems in the case where the interactions are long-range, and consequently, the macroscopic effective equations are described by non-linear fractional diffusion equations.}

\maketitle
	
\tableofcontents

\bigskip

\section{Introduction}

\subsection{Macroscopic Fluctuation Theory (MFT): Overwiew}

Macroscopic Fluctuation Theory (MFT) is the cornerstone of modern non-equilibrium statistical physics \cite{BDSGLJL15, M15, D07} that describes far from equilibrium processes and improves on Onsager's theory, in which fluctuations are modeled by Gaussian processes. It can be seen as an infinite-dimensional version of the Freidlin-Wentzel theory \cite{FW12}. One of the main interest of MFT is that it provides, in the context of interacting particle systems driven by external forces, a definition of a non-equilibrium free energy for the non-equilibrium stationary state (NESS)  as the solution of a dynamical variational problem. When the system is at equilibrium, the variational problem becomes trivial {\footnote{Of course, Gibbs equilibrium measures may have a very complicated form so that the variational problem becomes difficult to solve.}} in the sense that one recovers the usual equilibrium free energy given in the Gibbs formalism. 

Based on large deviations theory \cite{DV75, V84, E85, DS89, DH00, DZ09,T09}, MFT has been developed for interacting particle systems, mainly lattice gas, whose typical behaviour is given by a diffusion equation (e.g., Simple Symmetric Exclusion Process \cite{BDSGLJL03}), a viscous conservation law (e.g., Weakly Asymmetric Exclusion Process \cite{BD06, BLM09, BDSGLJL11}) or a conservation law (e.g., Asymmetric Simple Exclusion Process \cite{V04, M10, BBC18}). MFT has been extended to other lattice systems (e.g. \cite{BGL15}), to some reaction diffusion processes like Glauber--Kawasaki dynamics \cite{JLLV93, BL10} but also to systems with more than one conservation law \cite{B08, BKLL12,CRV14, V17} or to systems with ballistic transport \cite{DPSY23}. Without being exhaustive we mention other applications of the MFT like \cite{KMS15,MMS22}. It is also expected that it will be applicable to other fields like geophysics \cite{JL23} or turbulence \cite{R12}.

\subsection{Objectives}

The aim of this paper is to study the extension of the MFT framework for a certain class of lattice field models, known in the literature as Ginzburg-Landau (GL) dynamics. A convenient picture is to see them as stochastic fluctuating interfaces, see Figure \ref{Fig:interface}.  The novelty of the current paper w.r.t. the existing literature is that long-range interactions are considered. The hydrodynamic equation is then governed by a fractional diffusion equation \cite{V12}  (see Eq. \eqref{eq:hlGLG2}) instead  of a usual diffusion equation. Our main interest is to develop MFT for super-diffusive lattice field models and in particular to obtain some informations on the NESS of these open systems in the thermodynamic limit. 

\subsection{GL dynamics: Set-up}

To be more precize and in order to motivate the choice to consider GL dynamics,  let us first define them in a very general context, ignoring for the moment some external effects.  We are interested in the transport properties of a conservative lattice field model $\varphi_t :=\{ \varphi_t (x) \in {\mathcal M} \; ; \; x \in \Lambda\} \in {\mathcal M}^{\Lambda}$, where $\mathcal M$ is an arbitrary Riemannian sub-manifold of $\mathbb R^N$ and $\Lambda$ is a $D$-dimensional lattice. Depending on the physical context,  the variable $\varphi_t (x)$ represents the value at time $t$ of, e.g. a continuous scalar spin (${\mathcal M}=\mathbb R$), a charge (${\mathcal M}=\mathbb R$), an energy ($\mathcal M =\mathbb R^+$), a continuous N- spin model ($\mathcal M= S^{N-1} (1)$) etc.

A (lattice) $\varphi$-conservation law is locally expressed by an equation in the form 
\begin{equation}
\label{eq:conslaw1}
\partial_t \varphi_t (x) +\sum_{y \in \Lambda} J_t (x,y)=0, \quad x \in \Lambda \ , 
\end{equation}
where $J_t (x,y)$ is the instantaneous rate of the $\varphi$-current between site $x$ and site $y$ at time $t$, positively counted from $x$ to $y$, and such that $J_t(x,y) = -J_t (y,x)$.

The previous equation express that the change in time of $\varphi_t (x)$ is only due to exchanges between two arbitrary sites (there is no local creation or annihilation) so that, thanks to the antisymmetry of $J_t$, the following $\varphi$-conservation law holds:
\begin{equation}
\label{eq:conservationlawphi}
\partial_{t} \mathcal V_\Lambda (\varphi_t) =0  \quad \text{where} \quad    {\mathcal V}_\Lambda (\varphi) = \sum_{x \in \Lambda} \varphi (x)  \ .
\end{equation}
Note that  the number of conserved quantities is the same as the dimension of $\mathcal M$. 

A particular sub-class of $\varphi$-conservation law concerns (stochastic) lattice fields model where Eq. \eqref{eq:conslaw1} is a finite dimensional stochastic differential equation, i.e. whose the instantaneous rates of currents satisfy 
\begin{equation}
\label{eq:currentintro}
J_t (x,y) =-F_{\varphi_t} (x,y) - \sqrt{2 \Gamma_{\varphi_t} (x,y)} \ \dot\zeta_t (x,y), 
\end{equation}
where the deterministic vector field $F_{\varphi_t} (x,y)$ is a drift term, and $\dot\zeta_t (x,y)$ is a $N$-dimensional standard white Gaussian noise and the matricial field $\Gamma_{\varphi_t} (x,y)>0$ is the diffusivity{\footnote{Ito's convention is assumed everywhere in this article}}. To insure the antisymmetry of the current $J_t$, we have to impose that for any $x,y \in \Lambda$,
\begin{equation*}
F_{\varphi_t} (x,y) =- F_{\varphi_t} (y,x), \quad \dot\zeta_t (x,y)= - \dot\zeta_t (y,x), \quad \Gamma_{\varphi_t} (x,y)= \Gamma_{\varphi_t} (y,x) \ .
\end{equation*}
We also require that the white noises $\{\dot\zeta_t (x,y)\; ; \; x,y \in \Lambda\}$ are independent apart from the antisymmetry constraint{\footnote{Hence there are in fact only $|\Lambda| (|\Lambda| -1)/2$ independent $N$-dimensional white noises.}}. Injecting Eq. \eqref{eq:currentintro} in Eq. \eqref{eq:conslaw1} we obtain then a so-called {\textit{diffusion $\varphi$-conservation law}}. 

\bigskip
 \begin{remark} 
 Note that the main physical restrictions of this set-up are: 
 \begin{enumerate}[1.]
\item  The noise $\dot\zeta_t (x,y)$ associated to a link $\{x,y\}$  is a $N$-dimensional white Gaussian noise.  The physical origin of the noise is not fully justified. Relaxing the Gaussian and white hypothesis does not seem to be very considered until now in the probabilistic literature, despite that conservative lattice gas are peculiar case of $\varphi$-conservation law like in Eq. \eqref{eq:conslaw1}  with white Poissonnian noise. 
\item The instantaneous current  $J_t(x,y)$ is a function of the conservative field  $\varphi_t :=\{ \varphi_t (x) \in {\mathcal M} \; ; \; x \in \Lambda\}$.  There exist in the literature (see e.g. \cite{BBO06,LO}) plenty of $\varphi$-conservation law like Eq. \eqref{eq:conslaw1} where this is not the case and which therefore do not lead to a closed equation for the conservative field $\varphi$. 
\end{enumerate}
\medskip
Anyway, these two assumptions permit to ensure that the $\varphi$-conservation law Eq. \eqref{eq:conslaw1} defines a Markovian process. 
\end{remark}
 
\bigskip
Now, given a positive functional $\mathcal E_\Lambda (\varphi)$ called the GL Hamiltonian, our aim is to define a diffusion $\varphi$-conservation law of the previous form which is moreover invariant w.r.t. the Gibbs measure $d{\tilde \mu}^\Lambda_{0} (\varphi)\propto  e^{-\mathcal E_\Lambda (\varphi)} d \varphi$. We call such a dynamics a conservative GL dynamics.  If  $\Gamma_{\varphi} (x,y)$ is constant, say equal to one, it is sufficient to take the drift term associated to the gradient flow associated to $\mathcal E_\Lambda$ and respecting the $\varphi$-conservation law, i.e. $F_{\varphi}  (x,y) := -\partial_{\varphi (x)} \mathcal E_\Lambda + \partial_{\varphi(y)} \mathcal E_\Lambda$. However, as soon as $\Gamma_{\varphi}$ is not constant, the previous choice does not respect the fluctuation-dissipation relation and ${\tilde \mu}_{0}^\Lambda$ is no longer invariant. A possible choice to restore this discrepancy is to take the drift term $F_{\varphi}$  in the form\footnote{The fact that this choice implies that the Gibbs probability measure $d{\tilde \mu}^\Lambda_{0} (\varphi)\propto  e^{-\mathcal E_\Lambda  (\varphi)} d \varphi$ is an invariant measure of  the $\varphi$-conservation law,  will be proven in a peculiar set-up in the Section \ref{sec:model}.}  (see  \cite[Section 3.2.3.2]{RNCGPJS22} for more physical motivations in a slightly different context)
\begin{equation}
\label{eq:forceF}
\begin{split}
F_{\varphi} (x,y) 
& =- \Omega_{\varphi} (x,y) \ \left(\partial_{\varphi (x)} -\partial_{\varphi (y)} \right)   \  \mathcal E_\Lambda (\varphi) \ + \  \left(\partial_{\varphi (x)} -\partial_{\varphi (y)} \right)   \ \Omega_\varphi (x,y)  \ ,
\end{split}
\end{equation}
where for all $x,y \in \Lambda$, $\Omega_\varphi (x,y) = \Omega_\varphi (y,x)$ is a matricial field called mobility, and which satisfies the Einstein relation 
\begin{equation}
\label{eq:omega}
\Omega_ \varphi (x,y) + \Omega^\dagger_ \varphi (x,y) =2 \ \Gamma_{\varphi} (x,y) \ .
\end{equation}
 The anti-symmetric part of $\Gamma_\varphi$ can be of interest to model Hamiltonian (underdamped) effects.

Observe that since the dynamics is a $\varphi$-conservation law, we have more invariant  Gibbs measures apart from $d{\tilde \mu}_{0}^\Lambda (\varphi) \propto e^{-\mathcal E_\Lambda (\varphi)} d \varphi$:  for any chemical potential $\lambda \in \mathbb R${\footnote{Depending on the growth properties of $U$ we have sometimes to restrict the domain of the admissible $\lambda$'s in order to have well defined Gibbs measures. We will always assume that the dynamics is well defined and that the canonical Gibbs probability measures make sense, at least for a non-empty range of chemical potentials $\lambda$. Then $\mathbb R$ will have to be replaced by this domain.}},  the (canonical) Gibbs probability measure 
\begin{equation}
d \tilde \mu^\Lambda_\lambda (\varphi) \propto \exp \left(-\mathcal E_\Lambda (\varphi) -\lambda \mathcal V_\Lambda (\varphi)  \right) \ d\varphi 
\label{eq:FGM}
\end{equation}
is also invariant for the GL dynamics{\footnote{Hence the GL dynamics is not ergodic. But under suitable generic assumptions, one can prove it is, when the dynamics is restricted to a $\varphi$-invariant manifold $\left\{ \varphi \in {\mathcal M}^\Lambda \; ; \;  \mathcal V_\Lambda (\varphi) =C \right \}$, $C\in \mathbb R$ fixed. }}. In fact, if $\Gamma_\varphi=\Omega_\varphi$, then the canonical Gibbs measures $\tilde \mu^\Lambda_\lambda$ are reversible (i.e. satisfy the detailed balance condition).

\medskip
External interactions (boundary effects, external forces ...) can be incorporated to the free dynamics in several ways depending on the physical context of interest. We discuss few of them,  without being exhaustive. The free dynamics corresponds to the situation discussed above: the system is isolated from the exterior universe. If $\Lambda$ is the discrete $D$-dimensional torus (we say that we have periodic boundary conditions), we are are roughly in a similar physical situation. In both cases the canonical Gibbs measures  $\tilde\mu^\Lambda_\lambda$ are invariant for the dynamics. 

A first very different typical physical situation consists to add some $\varphi$-baths on the sites of the boundary $\partial\Lambda$ of $\Lambda$. The action of a bath acting on $x\in \partial\Lambda$ is to fix the distribution of $\varphi (x)$ equal to the $\varphi(x)$-marginal of $\tilde\mu^\Lambda_{\lambda (x)}$, where $\lambda(x)$ is an arbitrary fixed chemical potential depending on $x$. For example, we can model such a bath by a Langevin dynamics, acting on $\varphi (x)$ only, and making the $\varphi(x)$-marginal of $\tilde\mu^\Lambda_{\lambda (x)}$ reversible w.r.t. the Langevin dynamics. In this case, if the $\lambda (x)$ are non constant, none canonical Gibbs measure is invariant for the full dynamics{\footnote{If all the $\lambda (x)=\lambda$ are equal to a fixed value, under suitable generic conditions, we expect then that the unique invariant measure of the dynamics is $\tilde\mu^\Lambda_\lambda$.}}. We will be concerned with this last situation in the present article.

A second interesting situation (not discussed in the present paper) can result from the presence of some external forces acting in the bulk and not breaking the $\varphi$-conservation law. For example, transversal forces $F^{\perp}$ satisfying the antisymmetry property $F^{\perp}_{\varphi} (x,y) = - F^\perp_{\varphi} (y,x)$ and the transversal condition
\begin{equation}
\sum_{x,y \in \Lambda} \left(\partial_{\varphi (x)} -\partial_{\varphi (y)} \right)   \  \left( e^{-\mathcal E (\varphi)} \ F^\perp_\varphi (x,y) \right) = 0 
\label{eq:TF}
\end{equation}
do not break  the invariance of the canonical Gibbs measures. More generally, an external antisymmetric force $F^{ext}_{\varphi} (x,y) = - F^{ext}_{\varphi} (y,x)$ will still give rise to a diffusion $\varphi$-conservation law but will in general break the invariance of the canonical Gibbs measures{\footnote{We will investigate in fact a particular case of this situation in Section \ref{subsec:perturbed dynamics}.} \cite{FM88} . 

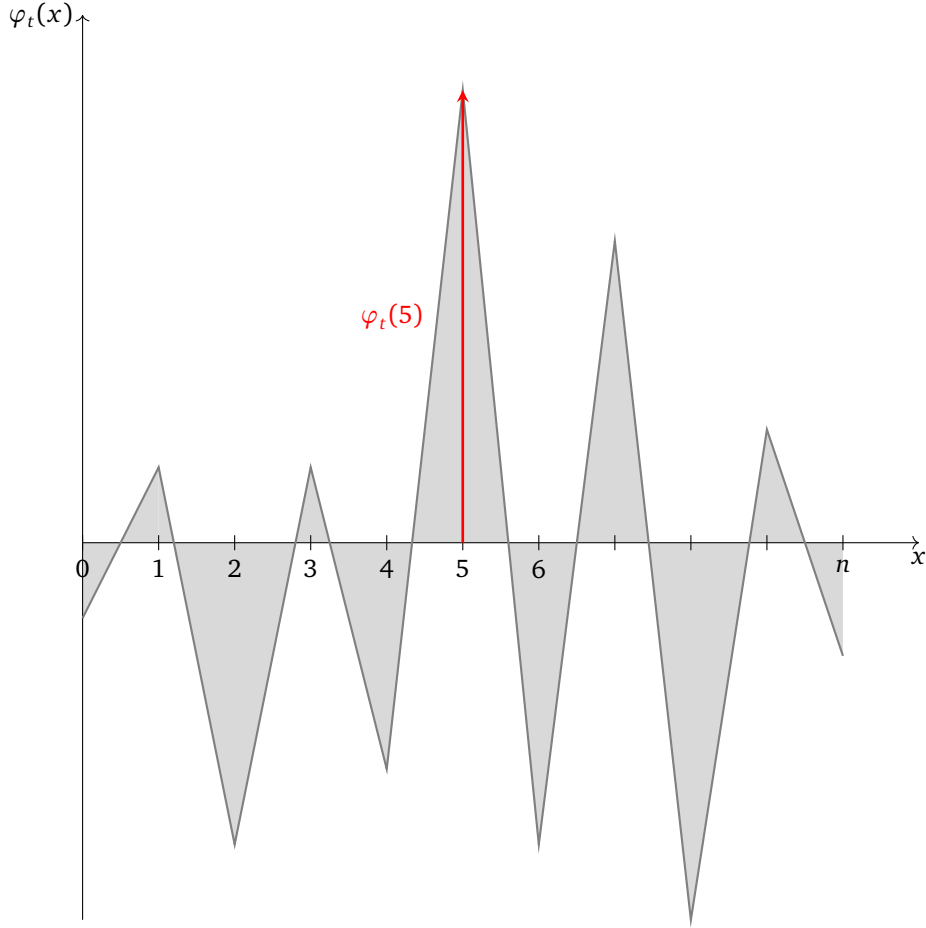
\begin{figure}
\centering
\label{Fig:interface}
\begin{tikzpicture}[scale=1]
\fill[gray!30] (0,0) -- (0,-1) -- (0.5,0) -- cycle;
\fill[gray!30] (0.5,0) -- (1,1) -- (1,0) -- cycle;
\fill[gray!30] (1,0) -- (1,1) -- (6/5,0) -- cycle;
\fill[gray!30] (6/5,0) -- (2,-4) -- (14/5,0) -- cycle;
\fill[gray!30] (14/5,0) -- (3,1) -- (3.25,0) -- cycle;
\fill[gray!30] (3.25,0) -- (4,-3) -- (13/3,0) -- cycle;
\fill[gray!30] (13/3,0) -- (5,6) -- (5.6,0) -- cycle;
\fill[gray!30] (5.6,0) -- (6,-4) -- (6.5,0) -- cycle;
\fill[gray!30] (6.5,0) -- (7,4)-- (67/9,0) -- cycle;
\fill[gray!30] (67/9,0) -- (8,-5) -- (114/13,0) -- cycle;
\fill[gray!30] (114/13,0) -- (9,1.5) -- (9.5,0) -- cycle;
\fill[gray!30] (9.5,0) -- (10,-1.5) -- (10,0) -- cycle;
\draw[->] (0,-5) -- (0,7) node[left] {{$\varphi_t (x)$}};
\draw[->] (0,0) -- (11,0) node[below] {{$x$}};

\foreach \x in {0,1,2,3,4,5,6}{
    \draw (\x,0.1) -- (\x,-0.1) node[below] {{$\x$}};
 }
\foreach \x in {7,8,9}{
    \draw (\x,0.1) -- (\x,-0.1) node[below]{$$};}       
\draw  (10,0.1) -- (10,-0.1) node[below] {{$n$}};   

\draw[gray, thick] (0,-1) -- (1,1) -- (2,-4) -- (3,1) -- (4,-3) -- (5,6) --(6,-4)-- (7,4) -- (8,-5) -- (9,1.5) -- (10,-1.5);

\draw[red, line width=1pt, >=stealth, ->] (5,0) -- (5,6);
\node[left] at (4.6,3) {\textcolor{red}{{$\varphi_t(5)$}}};
\end{tikzpicture}

\caption{The GL dynamics on the lattice $\Lambda_n=\{1,\ldots,n\}$ can be seen as a fluctuating interface conserving the algebraic volume (represented in gray) $\mathcal V_n (\varphi_t):= {\mathcal V}_{\Lambda_n} (\varphi_t)= \sum_{x \in \Lambda_n} \varphi_t (x)$ between the interface and the $x$-axis. } 

\end{figure}

\subsection{GL dynamics: Motivations and bibliography}

The discrete GL dynamics described in the previous paragraph can be physically motivated in many ways. 

\medskip 
{\underline{First}}: One goal \cite[Part II, Section 7.3]{Spohn} is to makes sense, via a discretisation, of some continuous ill-defined continuous GL dynamics, see e.g. \cite[Chap. 8.6]{CL95} or \cite{HH97}. In fact, eight (labeled by the letters A,B,C, ...,H) other ill-defined stochastic partial differential equations have been introduced in physics to model the dynamics of phase transitions. For example, the model A concerns a scalar continuous field, and is more or less an ad-hoc coarse grained version of Glauber Ising spin dynamics, hence without any conservation law. On the other hand, the model B is typically an ad-hoc scalar coarse grained version of Kawasaki spin dynamics model, and as one conservation law, and discretised version of it is is a case of our GL dynamics. The other models in this list concern vectorial continuous field will typically contain also non dissipative Hamiltonian forces. Discretised versions of theses models did not seem to have been considered until now in the physical mathematics literature. 

\medskip
{\underline{Second}}: They also appear as effective equations for the transport of energy in chains of oscillators with conservative bulk noise in a suitable weak coupling limit \cite{LO} or in weakly coupled Hamiltonian dynamics \cite{DL}. Even if \cite{LO, DL} are only concerned with short-range dynamics, understanding of energy transport problems in chains of oscillators with long-range interactions is the subject of recent several studies, e.g.  \cite{DCLLL19, BLL20, M20, WDX20, TS20, ALRT23}. 

\medskip
{\underline{Third}}: GL dynamics are standard models of interacting particle systems studied by probabilists since more than three decades and which has a rich history, see e.g. \cite{Spohn_1, F87, F87a, FM88, Fu89, F90, GPV, DV89, R90, Z90, V91, Y,  CY92, LY95, Q95, BLO97, GKRV09, BBP20} and references therein. Translated in our previous formalism,  all these studies are only concerned with the case $\mathcal M=\mathbb R$ or $\mathcal M=\mathbb R^+$ \cite{LO,DL} (but in any case $N=1$), so that $\Omega_{\varphi} (x,y) = \Gamma_{\varphi} (x,y)$ (see Eq. \eqref{eq:omega}) and with periodic boundary conditions (except{\footnote{The BEP model in \cite{GKRV09} is a GL dynamics but it is not mentioned in the paper.}} \cite{GKRV09}), so that $\Lambda:=\mathbb T_n$, the discrete torus of length $n$ (which will go to infinity is the scaling limit). In these works, the GL Hamiltonian $\mathcal E_{\mathbb T_n} (\varphi)$ is always supposed to be in the translation-invariance form
\begin{equation}
\label{eq:Evarphi}
\mathcal E_{\mathbb T_n} (\varphi) = \sum_{x \in \mathbb T_n} {\mathcal U} (\tau_x \varphi)
\end{equation}
where ${\mathcal U}$ is a function of $\varphi$ and the shift $\tau_x$ is defined via $(\tau_x \varphi)(z) =\varphi (z+x)$. Most of the previous papers consider the case where ${\mathcal U} (\varphi)=U(\varphi_0)$ for some given potential function $U$ (it will be our case also). Consistently, it is usually assumed that $\Gamma_{\varphi} (x,y)= \Gamma_{\tau_x \varphi}(0, y-x)$ and similarly for $\Omega_\varphi$. 
Moreover, as far as we know, \cite{Q95, B08a, GKRV09} are the only papers in the previous list to consider the case $\Gamma_\varphi$ non constant. However \cite{GKRV09} considers a reversible gradient system while \cite{Q95} considers reversible models which are non-gradient  (\cite{V91}, \cite[Part II, Section 2.4]{Spohn} \cite[Chapter 7]{KL99}, \cite{QY12}, \cite{BS22}). This means that the current in \cite{Q95} cannot be written as the discrete gradient of a local function. This adds very serious difficulties and \cite{Q95} has to rely on the famous Varadhan's non-gradient method. On the other hand \cite{B08a} considers a GL dynamics with a transverse drift and the model is non reversible{\footnote{Like in \cite{GKRV09} the author of \cite{B08a} did not notice that the dynamics introduced is a GL dynamics.}}. But, in fact, the most important remark for us is that in all these papers, only short-range interactions are considered in the sense that there exist $K, L \ge 1$ and $M \ge 0$ such that{\footnote{Usually, the authors take for convenience $K=1$, and apart from \cite{R90}, $M=0$.}}
\begin{equation}
\label{eq:intlocal}
\begin{cases}
&\Gamma_{\varphi} (x,y)=0 \ , \quad \text{if} \quad |x-y|>K \ ,\\
&\Gamma_\varphi (x,y) \quad \text{depends only of $\varphi (z)$ for $|z-x|\le L$ }\ ,  \quad  \text{if} \quad |x-y| \le K  \ ,\\
&\mathcal U \quad \text{ is local, i.e.  it depends only of $\varphi (z)$ for $|z|\le M$ \ . }
\end{cases}
\end{equation}
In these works are studied hydrodynamic limits \cite{F87,F87a, FM88, Fu89,  F90, GPV, R90, V91, Y, BLO97, GKRV09}, equilibrium and non-equilibrium fluctuations (i.e. Central Limit Theorems) \cite{Spohn_1, Z90, CY92} and dynamical large deviations  \cite{DV89, LY95, Q95, BBP20}. All of them are for isolated dynamics with periodic boundary conditions. The hydrodynamic limit obtained takes the form of a non-linear diffusion equations with periodic boundary conditions, apart from \cite{GKRV09} for which the diffusion equation is linear, with inhomogeneous boundary conditions, and \cite{BLO97} where is derived a `generalized' Cahn-Hillliard equation{\footnote{The assumptions done in the paper unfortunately rule out the  the obtention of the classical Cahn-Hilliard equation with quartic double well potential.}}.  The dynamical large deviations results for the empirical density associated to the field $\varphi$ appeared first in the seminal paper \cite{DV89} for a gradient Ginzburg-Landau dynamics defined on the one-dimensional discrete torus $\mathbb T_n$ of length $n$ going to infinity (with time rescaled diffusively by $n^2$). The result has been generalized in \cite{LY95} where the authors consider directly the dynamics in infinite volume, i.e. $\mathbb T_n$ is replaced by $\tfrac1n \mathbb Z$. A different approach to derive the result of \cite{DV89} is provided in \cite{BBP20}. In \cite{Q95}, Quastel derived a result similar to \cite{DV89} for $\Gamma_\varphi$ non constant and, as mentioned above, he has therefore to rely on the complex non-gradient methods and does not obtain an explicit form for the diffusion coefficient in the non-linear hydrodynamic equations. However, in all theses paper, the form of the dynamical large deviation function is the one considered in MFT for diffusive systems (see Eq. \eqref{eq:DLDF-USUAL}). 


\subsection{GL dynamics: Set-up  with long-range interactions}


The aim of this work is to study GL dynamics with long-range interactions when the lattice dimension is $D=1$ and the space dimension is $N=1$ (in fact we will restrict even our study to $\mathcal M=\mathbb R$ or $\mathcal M=\mathbb R^+$). There are several ways to introduce long-range effects in the GL dynamics since it suffices to break (at least) one of the conditions appearing in Eq. \eqref{eq:intlocal} by taking $K,L$ or $M$ infinite. We consider only the case where we break one of the three lines of Eq. \eqref{eq:intlocal}, while maintaining the other ones valid. Reading the existing literature, we think that the definition of long-range effects is very far to be uniform in the physical or mathematical literature. Hence, to be precize, for us, by definition, a long-range effect occurs as soon as, in its definition, the model investigated incorporates interactions which are not of finite-range. Observe that with this definition, long-range models can in fact behave like finite-range models at the macroscopic level and, more surprisingly, some short-range models can behave macroscopically like long-range models \cite{JKO15,BJG16}.  We describe below some possibilities while in the current paper we will discuss only one of them, i.e. the first one.\\
\begin{enumerate}[1)]
\item {\bf{Break of the validity of first line of Eq. \eqref{eq:intlocal}}}. This is the kind of long-range effects considered in this paper. We consider $\Gamma_\varphi (x,y)$ in the non-local form
\begin{equation*}
\Gamma_\varphi (x,y) = K(y-x) \beta (\varphi(x), \varphi (y))
\end{equation*}
where $\beta$ is a smooth positive function on $\mathbb R^2$ and $K$ is a long-range coupling given by
\begin{equation}
\label{eq:KLR}
K(z)=\frac{\sc 1_{z \ne 0} }{|z|^{1+\gamma}}
\end{equation}
where $\gamma>0$. Moreover, we assume that the GL Hamiltonian $\mathcal E_\Lambda$ is in the form \eqref{eq:Evarphi} but with a local function $\mathcal U$, i.e. $\mathcal U (\varphi)$ depends only on the variables $\varphi_z$ for $|z| \le M$, $M$ fixed. For example
\begin{equation}
\label{eq:Ephisimple}
\mathcal E_{\Lambda} (\varphi)= \sum_x {\mathcal U} (\tau_x \varphi)= \sum_{x} U(\varphi (x)) 
\end{equation}
where $U$ is a given nice{\footnote{Recall that we will always assume that the dynamics are well defined and that the Gibbs canonical measures make sense, at least for some non-empty range of chemical potentials $\lambda$.}} potential function, i.e. $\mathcal U (\varphi)=U(\varphi(0))$. 

For lattice gas, this kind of long-range interactions have been investigated in \cite{J08, BJ17 , SS18, BGJ19, BGJ21,GS22, BCGS} and is restricted to hydrodynamic limits which are linear (apart from \cite{J08,SS18}).\\

\item {\bf{Break of the validity of the second line of Eq. \eqref{eq:intlocal}}}. While this possibility would make sense we are note aware of any result in this direction (by breaking moreover or not the first line of Eq. \eqref{eq:intlocal}).\\

\item {\bf{Break of the validity of the third line of Eq. \eqref{eq:intlocal}}}. In this case we maintain thus a finite range $\Gamma$ in the sense of the two first lines of the definition \eqref{eq:intlocal}, but introduce long-range interactions in the GL Hamiltonian $\mathcal E$ by assuming the function $\mathcal U$ is non local. Typically (the reader will generalise easily) we have still Eq. \eqref{eq:Evarphi}, but now $\mathcal U$ is in the form 
\begin{equation*}
\mathcal{U}\left(\varphi\right)
= U\left(\varphi(0)\right) + \sum_{z\in\Lambda} K (z) \ V\left(\varphi(0),\varphi(z)\right)
\end{equation*}
where $U$ is the one-site potential, $V$ is the interaction potential and $K$ the coupling function. As usual we assume $U$ and $V$ are so that the dynamics are well defined and that the Gibbs canonical measures make sense, at least for some non-empty range of chemical potentials. In particular we do not consider singular potentials and we assume that $U \ge 0$ grows sufficiently fast to infinity in order to control also the growth of $V$.

There are then several possibilities to introduce long-range effects (in $K$, in $V$ or in both): \\
\begin{enumerate}[a)]
\item For the first possibility we take $K$ a function like in Eq. \eqref{eq:KLR} and $V$ smooth, e.g. a polynomial function. A quite similar case to this has been considered in \cite{Y94}. There Yau considers a GL Hamiltonian in the Kac's form
\begin{equation*}
\mathcal E_{\mathbb T_n}  (\varphi)= \sum_{x \in \mathbb T_n} U(\varphi (x)) \ + \ \sum_{x,y \in \mathbb T_n} n^{-a} J \left(\tfrac{y-x}{n^a} \right) \varphi (x) \varphi (y) 
\end{equation*}
with $0<a<1$ and $J$ a non-negative function with compact support such that $\int_{\mathbb R} du \ J(u)  =1$. We recall that $\mathbb T_n$ is the discrete torus of length $n$ and that in the scaling limit, $n$ goes to infinity. Hence, here, $K(z)=n^{-a} J(z/n^a)$ is long-range and $V(\phi, \phi') = \phi \phi'$ is short-range.  The hydrodynamic equation is still given by a nonlinear diffusion equation like in the short-range case since the `long-range' effects in \cite{Y94} are in fact sufficiently weak. This kind of long-range effects have been lengthly studied in the context of lattice gas (without being exhaustive, we refer the reader to  \cite{G91, LOP91, GL97, AG98, GL98, GLM00, DMPT11, MO12}).\\

\item We take $K$ with finite range but $V$ with long-range support. While it could make sense we are not aware of any result in this direction. \\

\item A third possibility is to take $K$ long-range and $V$ long-range. We can for example consider $\mathcal U (\varphi)$ in the form  
\begin{equation*}
{\mathcal U} (\varphi) =U(\varphi(0)) \ + \ \sum_{z} V (\varphi (0), \varphi(z)) \ , 
\end{equation*}
where $V:\mathbb R \to \mathbb R$ is long-range, i.e. $V(\phi, \phi')= 1/{|\phi -\phi'|^{1+\gamma}}$, $\gamma>0$. It is two times long-range in the sense that $K(z):=1$ does not have even any decay as $|z| \to \infty$. Apparently this case has not be investigated in the literature for GL dynamics.  Of course, this possibility can not provide long-range effects for lattice gas since the occupation variables $\varphi (0), \varphi (z)$ take values in $\{0,1\}$.

\end{enumerate}

\end{enumerate}

\bigskip
As mentioned above, in this paper, we will be only concerned with the situation 1) with $\gamma<2$. If $\gamma>2$ we expect to have a diffusive behaviour, but the treatment of this case would require, apart from specific cases, to develop non-gradient tools like in the short range case \cite{Q95}. This is out of the scope of the present paper (see Remark \ref{rem:gamma2}). 

\bigskip
\begin{remark}
\label{rem:QF}
To conclude this bibliographical interlude, let us mention there exists also an important literature for interacting Brownian particles \cite[Part II, Section 7.2]{Spohn} with long range interactions (also called sometimes mean-field models)  \cite{DZ78, D83, SK86, DG87, G88, D96, BL99,BDF12, BGN16,BBCCM20} and an exponentially increasing interest for related models motivated by active matter \cite{VCBJSS95,BM08,PDB08,BCMP15, CDP18}. Of course, a system of interacting Brownian particles (or models in  active matter) can be seen as lattice field models if one labels the particles and defines $\varphi_t (x)$ as the position{\footnote{For models in active matter, it would be also necessary to incorporate extra variables like velocity or angles.}} of the particle with label $x$ at time $t$. However they do not provide a $\varphi$-conservation law. The conserved quantity of interest in these models is the particle number so that the interpretation as lattice models is not really interesting{\footnote{This is different in the context of lattice gas where typically there, for $x$ in $\Lambda$ and $\eta$ a given configuration, $\eta (x) \in \{0,1\}$ denotes the number of particles on site $x$. A (conservative) lattice gas is an $\eta$-conservation law so that the analogy with GL dynamics is clear when we replace $\eta$ by $\varphi$.}}. In \cite{LD22, FLDMS23} the authors study properties of the stationary state of a one dimensional Lieb-Liniger delta Bose gas.  In \cite{DKM23} the authors developed MFT for a system composed of Brownian motions on the line interacting through the long range Riesz potential. The authors then obtained there the long time behaviour of the variance of the fluctuations of the integrated current and of the position of a tagged particle. Connected to these models exists also some literature motivated by active matter. In particular, let us mention \cite{TLDS23a, TLDS23} where is studied a Dyson Brownian motions model  in which run and tumble particles interact via a logarithmic repulsive potential in the presence of a harmonic well. The hydrodynamic limit of the systems there is then governed by a nonlinear fractional diffusion equation{\footnote{In \cite{DKM23}, if $s>1$, the effective dynamics is in fact short-range and the hydrodynamics is given by a standard diffusion equation.}} but with a quadratic flux. The obtention of a quadratic flux will be a very particular case for us.  Let us also mention that in these models particles evolve on the infinite line and that the system is not subject to boundary thermal forces (without external forces in \cite{LD22, DKM23, FLDMS23} and with external ones in \cite{TLDS23a,TLDS23}).
\end{remark}

\bigskip

\bigskip
\subsection{Plan of the paper}


After having introduced the model in Section \ref{sec:model} we derive the hydrodynamic and hydrostatic limits of the model in Section \ref{sec;hlhs} (see Eq. \eqref{eq:hlGLG2} and Eq. \eqref{eq:hlGLG-ss}). The hydrodynamic limit is derived in a subdiffusive time scale, i.e. shorter than the usual diffusive time. This is done for the original model as well as for a perturbed dynamics (see Section \ref{subsec:perturbed dynamics}) used later to establish a large deviation principle. In Section \ref{sec:dynamical LD} we derive the (dynamical) large deviations functional of the empirical density of the volume in the previous subdiffusive time scale: the probability to observe an atypical macroscopic profile, i.e. different from the one provided by the solution of the hydrodynamic equation, is exponentially small in the system size, and we identify the corresponding rate (see Eq. \eqref{eq:expressionLDF2}). In Section \ref{sec:MFT} we identify the non-equilibrium free energy (or quasi-potential) of the NESS as the solution of a stationary Hamilton-Jacobi equation (see Eq. \eqref{eq:HJ-equation-einstein}). In Section \ref{sec:difflimit} we study our results in the diffusive limit $\gamma \to 2^-$ and argue the existence of a $0$-th order phase transition. Section \ref{sec:particular-case} considers special cases. In Section \ref{sec:comparison} we investigate the differences and similarities between the diffusive case (nearest neighbour interactions) and the superdiffusive case ( heavy tails long range interactions).  The paper is completed by several appendices. In a companion paper \cite{BCK24} we will extend the results of this paper to a more general set-up including in particular  lattice gas with long-range interactions, and revisit the arguments in the framework of fluctuating hydrodynamics.


\bigskip 

\section{Long-range GL dynamics}
\label{sec:model}
Let $\gamma<2$ and let $K:\mathbb R \to (0,\infty]$ be the function defined by 
\begin{equation}
\label{eq:defK}
K(u)= \frac{1}{\vert u \vert^{1+\gamma}} \;\; \text{if}\;\;  u\ne0, \quad K(0)=0 \ .
\end{equation}

\bigskip
We restrict our general set-up of the introduction to the case where the lattice  $\Lambda$, now denoted $\Lambda_n=\{1,\ldots, n\}$, $n\ge 1$,  is of dimension $D=1$ and the $\varphi$'s-space $\mathcal M$ is equal to  to $ \mathbb R$ or $\mathbb R^+$ (hence $N=1$). We also assume the GL Hamiltonian is without interactions, i.e.  given by Eq. \eqref{eq:Ephisimple} with  $U$ a smooth potential 
\begin{equation}
\mathcal E (\varphi) := \mathcal E_{\Lambda_n} (\varphi) =\sum_{x \in \Lambda_n} U (\varphi (x)) \ .
\label{eq:E}
\end{equation}
Finally, we assume also the choice
 \begin{equation}
\label{eq:fatigue}
\Gamma_\varphi (x,y) := K(y-x) \beta (\varphi(x), \varphi (y)) \ , 
\end{equation}
 in Eq. \eqref{eq:forceF}, with $\beta (\varphi (x) , \varphi(y))$ is a symmetric positive function. By Eq. \eqref{eq:forceF} and the one-dimensional version of Eq. \eqref{eq:omega},  the drift term $F_\varphi$ is then
\begin{equation}
\label{eq:alpha-gamma}
\begin{split}
&F_{\varphi} (x,y):= K (y-x) \alpha (\varphi (x), \varphi(y))\\
&=-K(y-x) \beta\left(\varphi(x),\varphi(y)\right)\left(U'\left(\varphi(x)\right)-U'\left(\varphi(y)\right)\right)+ K(y-x)\left(\partial_{\varphi(x)}-\partial_{\varphi(y)}\right)\beta \left(\varphi(x),\varphi(y)\right)\\
&=K(y-x) e^{\mathcal E (\varphi)} (\partial_{\varphi(x)} -\partial_{\varphi(y)}) \left( e^{-\mathcal E(\varphi)} \beta (\varphi (x), \varphi (y))  \right) 
\end{split}
\end{equation}
which is an antisymmetric function in $\varphi(x),\varphi(y)$.


This choice is mainly motivated by the fact there are the long-range version of the models appearing in \cite{V91, Q95, LO, DL}.

\bigskip

\subsection{Long-range GL dynamics with free boundary conditions}
 
\

Before to introduce the GL dynamics with long-range interactions in contact with baths, let us consider the closed dynamics with free boundary conditions. 
A typical configuration of a Ginzburg-Landau dynamics is denoted by $\varphi =\{ \varphi (x) \in \mathbb R \; ; \; x \in \Lambda_n\}$ and the configuration space is thus $\mathbb R^{\Lambda_n}$. 


The equations of motion defined through Eqs. \eqref{eq:conslaw1}, \eqref{eq:currentintro}, \eqref{eq:fatigue} and \eqref{eq:alpha-gamma},  are given for any $x \in \Lambda_n$ by the Ito's stochastic differential equation
\begin{equation}
\partial_t \varphi_t (x) = \sum_{y \in \Lambda_n} K(y-x) \alpha (\varphi_t (x), \varphi_t (y)) +\sqrt{2} \sum_{y \in \Lambda_n} \sqrt{ K (y-x) \beta(\varphi_t (x), \varphi_t (y))}  \ \dot \zeta_t (x,y) \ .
\label{eq:sdeB}
\end{equation}
Here, we recall that $\{ \dot \zeta_t (x,y) \; ; \; x<y\}$ are independent standard white noises and $\dot \zeta (y,x) = - \dot \zeta (x,y)$ (by convention $\dot \zeta (x,x) =0$).

The Gibbs probability measure{\footnote{To lighten the notations the dependence in $\Lambda_n$ is omitted.}} $d{\tilde \mu}_0 (\varphi) \propto  e^{-\mathcal E (\varphi)} d \varphi$ is a reversible, hence invariant, measure for this dynamics.  This comes from the form of the Markovian generator $\mathcal G_b^n$ of the dynamics  \eqref{eq:sdeB}
 (see Appendix \ref{app:bulkG} for a proof)
\begin{equation}
\label{eq:Gb}
\begin{split}
\mathcal G_b^n &= \frac{1}{2}\sum_{x,y \in \Lambda_n} K (y-x)\  e^{\mathcal E (\varphi)} (\partial_{\varphi (y)} -\partial_{\varphi (x)}) \left[ e^{- \mathcal E (\varphi)} \  \beta (\varphi(x), \varphi(y)) \ (\partial_{\varphi (y)} -\partial_{\varphi (x)} ) \right] \ .
\end{split}
\end{equation}
Moreover, since the volume $\mathcal V (\varphi):= \mathcal V_{\Lambda_n} (\varphi)$ is a conserved quantity of the dynamics, the Gibbs probability measures ${\tilde \mu}_\lambda:={\tilde \mu}_{\lambda}^{\Lambda_n}$ (see Eq. \eqref{eq:FGM}) parameterised by $\lambda \in \mathbb R${\footnote{Depending on the growth properties of $U$ we have sometimes to restrict the domain of the admissible $\lambda$'s in order to have well defined Gibbs measures.}} and given by
\begin{equation}
\label{eq:GMF2}
{\tilde \mu}_\lambda (d\varphi) = {\mathcal Z}_\lambda^{-1} \exp\left( -\mathcal E (\varphi) - \lambda \mathcal V (\varphi) \right) d\varphi \ ,
\end{equation}
forms the family of extremal{\footnote{This is a consequence of Appendix \ref{app:ergodicity}.}} product{\footnote{The fact that the invariant measure is  produced by i.i.d. random variables does not mean that there are no interactions in the dynamics: the stochastic differential equations appearing in Eq.  \eqref{eq:sdeB} igive rise to an interacting diffusion process, even when $\beta=1$. }}  invariant probability measures of the dynamics. Here $\mathcal Z_\lambda = (Z(\lambda))^{n}$, where 
$$Z(\lambda)=\int d \phi\exp\left(-U\left(\phi\right)-\lambda\phi\right)$$
is the partition function. In fact,  Eq. \eqref{eq:reversibility} below shows that the dynamics satisfies detailed balance condition with respect to $\tilde\mu_\lambda$. 

Since the quantity of interest is the volume it is more convenient to perform a change of parameterization in the equilibrium measures. We denote by $\nu_\Phi$ the probability measure on $\mathbb R$ such that
\begin{equation}
\label{eq:nu_Phi}
\nu_\Phi(d\phi)= Z^{-1} (\lambda) \ \exp\left( -U(\phi)- \lambda \phi \right) \ d\phi 
\end{equation}
with $ Z(\lambda)$ the normalisation constant and $\lambda:=\lambda(\Phi)$ chosen such that $\langle \phi\rangle_{\nu_\Phi} = \Phi$. Then the product probability measures
\begin{equation}
\label{eq:muphidef}
    \mu_{\Phi}:={\tilde \mu}_{\lambda(\Phi)}
\end{equation}
obtained in this way are the canonical equilibrium measures of the dynamics.

\bigskip
\begin{remark}
In this one-dimensional set-up, the $\varphi$-conservation law given by Eq. \eqref{eq:conslaw1} or Eq. \eqref{eq:sdeB} can be rewritten in the form of the discrete continuity equation 
 \begin{equation*}
\partial_t \varphi_t (x) \  + \ J_t (x)- J_t(x-1)=0 \ .
\end{equation*}
where $J_t (x)$ is the instantaneous local rate of $\varphi$-current at $x$, expressed in terms of the instantaneous rate  of $\varphi$-currents
\begin{equation}
\label{eq:Jcurrent}
J_t (x) =\sum_{y \le x} \sum_{z \in \Lambda_n}  J_{t}(y,z) \ .
 \end{equation}
Then, with the relations provided in Eqs. \eqref{eq:currentintro},  \eqref{eq:fatigue} and \eqref{eq:alpha-gamma}, we obtain 
\begin{equation}
\label{eq:ic}
J_t (x) = -\sum_{y \le x} \sum_{z \in \Lambda_n} \left\{ K (z-y) \alpha (\varphi_t (y),\varphi_t (z)) + \sqrt{K (z-y) \beta (\varphi_t (y), \varphi_t (z))} \ \dot\zeta_t (y,z)\right\}.
\end{equation}
 \end{remark}

 \begin{remark}
 The  conserved quantity 
 \begin{equation}
 \label{eq:V}
 \mathcal V (\varphi_{t}) := \sum_{x\in \Lambda_n} \varphi_{t}(x)
 \end{equation}
 defined in Eq.  \eqref{eq:conservationlawphi} is sometimes called in the following the `volume', and it is in fact the only conserved quantity (see Appendix \ref{app:ergodicity} for a proof). Moreover, if we restrict the dynamics to the sub-manifold (hyperplane) $E_C=\{ \varphi \in {\mathbb R}^{\Lambda_n}\; ;\; \mathcal V (\varphi) =C \}$, where $C$ is an arbitrary constant, then the dynamics is ergodic and its unique invariant measure is the uniform measure on $E_C$ (see Appendix \ref{app:ergodicity} for a proof). In other words the microcanonical measures of the dynamics with free boundaries are given by the uniform probability measures on $(E_C)_{C \in \mathbb R}$. 
  \end{remark}
 
 \bigskip 
 \begin{remark}
In the GL dynamics derived in \cite{LO,DL}, in the context of transport of energy in chains of oscillators,  we have that
\begin{equation*}
\beta (\phi,\phi')=\phi \phi' \Theta (\phi,\phi') 
\end{equation*}
where $\Theta$ is a smooth positive function. It is then easy to check that in this case, if we start the dynamics with an initial condition $\varphi_0$ such that $\varphi_0 (x) >0$ for any $x$, then at any time $t\ge 0$, $\varphi_t (x) >0$ for any $x$. This is consistent with the fact that the variable $\varphi_t (x)$ is the value of the energy (which is positive) in the effective dynamics derived by a suitable scaling limit in \cite{LO,DL}.
\end{remark}

\subsection{GL thermodynamical quantities}

Since the Gibbs equilibrium measures are product, see Eq. \eqref{eq:GMF2}, the corresponding thermodynamic relations are reduced to the study of the ones for the marginal $\nu_{\Phi}$. By introducing the (concave{\footnote{It follows from H\"older's inequality.}}) equilibrium free energy 
\begin{equation}
\label{eq:Free}
F (\lambda)= -\log Z(\lambda) 
\end{equation}
and the thermodynamical (microcanonical) convex{\footnote{The supremum over a family of affine functions is convex.}} entropy function
\begin{equation}
\label{eq:entropy-function}
S(\Phi)= \sup_{\lambda} \left\{F(\lambda) -\lambda \Phi  \right\}
\end{equation}
we get that, if $F(\lambda)$ is differentiable,  the usual Legendre duality relations hold
\begin{equation}
\label{eq:dualityLegendre}
\lambda (\Phi) = -S' (\Phi) , \quad \Phi =F' (\lambda (\Phi))    \ .
\end{equation}
Moreover, if $F''$ exists and is fintie, we also have the static fluctuation dissipation theorem \cite{K66,KTH95,LB04}:
\begin{equation}
\label{eq:cov:equilibrium}
- F'' (\lambda (\Phi))  = \left\langle \phi^2 \right\rangle_{\nu_\Phi} - (\left\langle \phi \right \rangle_{\nu_\Phi})^2 = \cfrac{1}{S'' (\Phi)} \ge 0\ ,
\end{equation}
where the last equality follows from \eqref{eq:dualityLegendre}. The last inequality shows then that the entropy function $S (\Phi)$ is strictly convex, i.e. 
\begin{equation}
\label{eq:Conv}
S'' (\Phi) > 0 \ .
\end{equation}
 }

\bigskip
\begin{remark}
For the reader not very familiar with the physicist's jargon, let us point out that the equilibrium free energy $F$ and the entropy $S$ are just standard probabilistic objects appearing in the large deviations theory. More precisely, we have the relations
\[
F\left(\lambda\right)=F(0)-\Lambda_{\widetilde{\mu}_{0}}(-\lambda)\ , \quad S(\Phi)=F(0)+I_{\widetilde{\mu}_{0}}(\Phi) \ , 
\]
where $\Lambda_{\widetilde{\mu}_{0}}$ (resp. $I_{\widetilde{\mu}_{0}}$
) is the scaled cumulant generating function (resp. the large deviation
rate function) of the conserved quantity \label{eq:V} $\mathcal V_{{\Lambda_n}}  (\varphi):=\sum_{x\in\Lambda_{n}}\varphi(x)$
under the product probability measure ${\tilde \mu}^{\Lambda_n}_0 (d\varphi) \propto  \exp\left( -\mathcal E_{\Lambda_n}  (\varphi)  \right) d\varphi$,  defined formally
as  \cite{DV75, V84, E85, DS89, DH00, DZ09,T09}
\begin{equation}
\begin{split}
\Lambda_{\widetilde{\mu}_{0}}(\lambda) &:= \lim_{n\rightarrow\infty} n^{-1} \log\left\langle  e^{\lambda \mathcal V_{\Lambda_n} (\varphi)}\right\rangle_{\widetilde{\mu}_{0}} \ , \\
I_{\widetilde{\mu}_{0}}(\Phi)& := -\lim_{n\rightarrow\infty} n^{-1} \log  \left\langle \delta\left(\mathcal V_{\Lambda_n}(\varphi)-n\Phi\right) \right\rangle_{\widetilde{\mu}_{0}}  \ .
\end{split}
\end{equation}
With this in mind, the relation of Eq.\eqref{eq:entropy-function} is a particular case of Ellis-G\"artner Theorem  \cite{DV75, V84, E85, DS89, DH00, DZ09,T09},  valid as soon as  $F$ exists, modulo some more technical assumptions apart from differentiability. We must also point that the entropy in the physical literature is most of the time defined as $-S$, and is then concave. We did not choose this convention here in order to avoid many negative signs in several formula. 
\end{remark}

\subsection{Long-range GL dynamics in contact with two baths}


At the boundaries we introduce some baths, which will break the detailed balance condition in the steady state, and which will be responsible for a non vanishing macroscopic volume flux. Fix two constants $\Phi_\ell$ and $\Phi_r$. Then the equations of motion for the particle $1$ and particle $n$ are now
\begin{equation}
\begin{split}
\partial_t \varphi_t (1) &= -( \lambda(\Phi_\ell) + U' (\varphi_t (1))) + \sqrt{2} \dot\zeta_t (0,1) \\
& \ +\ \sum_{y \in \Lambda_n} K (y-1) \alpha (\varphi_t (1), \varphi_t (y)) +\sqrt{2} \  \sum_{y \in \Lambda_n} {\sqrt{K (y-1) \beta (\varphi_t (1), \varphi_t (y))}}  \ \dot \zeta_t {(x,y)}\ , \\
\partial_t \varphi_t (n) &= -(\lambda(\Phi_r) +U'(\varphi_t (n))) + \sqrt{2} \ \dot\zeta_t {(n,n+1)} \\
& \ +\ \sum_{y \in \Lambda_n} K (y-n) \alpha (\varphi_t (n), \varphi_t(y))  \ +\sqrt{2} \sum_{y \in \Lambda_n} \sqrt{K (y-n) \beta(\varphi_t (n), \varphi_t (y))}  \ \dot \zeta_t {(x,y)}\ , 
 \end{split}
 \label{eq:Bord}
\end{equation}
where $\dot\zeta {(0,1)}, \dot\zeta{(n,n+1)}$ are two new independent standard white noises. The two quantities $ \lambda(\Phi_\ell), \lambda (\Phi_r)$ have been defined previously by Eq. \eqref{eq:dualityLegendre}. 

\bigskip 
Let us comment about the choice of the form of the baths. Consider for example the bath on the left. We want to define a dynamics at the left boundary (site $1$) such that the invariant measure of this dynamics satisfies the detailed balance condition with respect to $\nu_{\Phi_\ell}$ (recall Eq. \eqref{eq:nu_Phi}), since we want to fix the value of the volume at the left boundary to be $\Phi_\ell$. The dynamics introduced above is thus the most natural choice.  

In the presence of the baths, the volume $\mathcal V (\varphi_{t})$ defining in Eq. \eqref{eq:conservationlawphi}  is no longer conserved but,  if $\Phi:=\Phi_\ell =\Phi_r$, then the systems is still at equilibrium: it satisfies the detailed balance condition with respect to $\mu_{\Phi}={\tilde \mu}_{\lambda(\Phi)}$.

\bigskip

The Markovian generator of the open dynamics is now given by 
\begin{equation*}
\mathcal G^n =\mathcal G_\ell +\mathcal G_r +\mathcal G^n_b \ ,
\end{equation*}
where $\mathcal G_b^n $ is the generator of the dynamics with free boundary conditions given in Eq. \eqref{eq:Gb} and
\begin{equation}
\label{eq:Gbo}
\mathcal G_\ell = -(\lambda(\Phi_\ell) +U'(\varphi (1)))  \partial_{\varphi(1)} + \partial^2_{\varphi(1)}, \quad  \mathcal G_r = -(\lambda(\Phi_r) + U' (\varphi (n)))  \partial_{\varphi(n)} + \partial^2_{\varphi(n)} 
\end{equation}
are the generators of the baths.

\bigskip 

\begin{remark}
In other words, the Fokker-Planck equation of the GL dynamics described above takes the form 
\begin{equation*}
\partial_t P_t (\varphi) = {\mathcal L}^n P_t (\varphi)
\end{equation*}
where $\mathcal L^n$ is given by
\begin{equation*}
\mathcal L^n =\mathcal L_\ell +\mathcal L_r +\mathcal L^n_b \ .
\end{equation*}
The bulk part is given by
\begin{equation}
\label{eq:Lb}
\begin{split}
\mathcal L_b^n &=  \frac{1}{2}  \sum_{x,y \in \Lambda_n} K(y-x)\  (\partial_{\varphi (y)} -\partial_{\varphi (x)}) \left[  e^{- \mathcal E (\varphi)}\  \beta (\varphi(x), \varphi(y)) \ (\partial_{\varphi (y)} -\partial_{\varphi (x)} ) \ e^{\mathcal E (\varphi)} \right] \\
\end{split}
\end{equation}
and the boundary parts by 
\begin{equation}
\label{eq:Lbo}
\mathcal L_\ell = \partial^2_{\varphi(1)} + \partial_{\varphi(1)} (U'(\varphi (1)) +\lambda(\Phi_\ell) ), \quad  \mathcal L_r = \partial^2_{\varphi(n)} + \partial_{\varphi(n-1)} (U'(\varphi (n)) +\lambda(\Phi_r) )\ .
\end{equation}

\bigskip

In particular, with this expression, it is easy to prove that the dynamics with free boundary conditions satisfies the detailed balance condition with respect to $\tilde \mu_\lambda$ for any $\lambda$ since for any function $f$, we have
\begin{equation}
\label{eq:reversibility}
\widetilde{\mu}_{\lambda}(\varphi)\mathcal{G}_{b}^{n}\left(\frac{f(\varphi)}{\widetilde{\mu}_{\lambda}(\varphi)}\right)=\mathcal{L}_{b}^{n}\left(f(\varphi)\right) \ .
\end{equation}
\end{remark}

\bigskip

\section{Hydrodynamic and hydrostatic limits}
\label{sec;hlhs}

\subsection{Hydrodynamic limit}

In this section, we will prove that the hydrodynamic limit is given by the nonlinear fractional{\footnote{The word fractional is justified by the fact the operator $\mathcal A$ can be seen as a generalisation of the classical linear fractional Laplacian \cite{V12}.}} diffusion equation with Dirichlet boundary condition
\begin{equation}
\label{eq:hlGLG2}
\begin{cases}
&\partial_t \Phi_t (u) - {\mathcal A} (\Phi_t) \ (u) =0 \ ,\\
&\Phi_t (0)=\Phi_\ell, \quad \Phi_t (1)= \Phi_r \ ,\\
&\Phi_t \vert_{t=0} = \Phi_0 \ ,
\end{cases}
\end{equation}
where the nonlinear operator $\mathcal A$ is defined by 
\begin{equation}
\label{eq:def-A2}
\begin{split}
 {\mathcal A} (\Phi_t) \ (u) &:= \int_0^1 dv \ K(v-u) \ A (\Phi_t (u), \Phi_t (v))\  , \end{split}
\end{equation}
with 
\begin{equation}
A(\Phi,\Phi') = \int d\nu_\Phi (\phi)d\nu_{\Phi'} (\phi')  \ \alpha (\phi,\phi') \ .
\label{eq:A2}
\end{equation}

\bigskip 

\begin{remark}
Note that this equation depends on the `microscopic'  quantities $K,U,\beta$, present in the microscopic SDE defined by Eq. \eqref{eq:sdeB}, but also on the `emergent' quantity $S(\Phi)$ (which appears in the measure $\nu_\Phi$ given by Eq. \eqref{eq:nu_Phi}).
\end{remark}

\bigskip

\begin{remark}
We can rewrite 
\begin{equation}
\label{eq:def-A-2}
\begin{split}
 {\mathcal A} (\Phi_t) \ (u) &:= \partial_u \left( \int_0^u dv\  \int_{0}^1 dw \ K(w-v) \ A \left( \Phi_{t} (v), \Phi_{t} (w) \right)\right) 
\end{split}
\end{equation}
and the term 
\begin{equation}
\label{eq:current-form}
J(\Phi (u)) =-\int_0^u dv\  \int_{0}^1 dw \ K(w-v) \ A \left( \Phi (v), \Phi (w) \right) = -\int_0^u dw \ \mathcal A (\Phi) (w) 
\end{equation}
is then the macroscopic instantaneous current associated to the nonlinear fractional diffusion appearing in Eq. \eqref{eq:hlGLG2}. As mentioned in the Remark \ref{rem:QF} of the Introduction, this current is usually nonlinear and even not necessarily quadratic.
\end{remark}

\hspace{1cm}

\begin{proof}
Since $\gamma <2$, we have to look at the dynamics in a subdiffusive time scale, i.e. a time scale shorter than the diffusive one. Therefore we define the empirical volume profile at time $t n^\gamma$ by
\begin{equation}
\pi^n_{tn^\gamma}(u)=\cfrac1n\sum_{x \in \Lambda_n} \varphi_{tn^\gamma} (x) \delta(u-\tfrac xn), \quad u\in [0,1] \ .
\label{eq:ev}
\end{equation}
For any smooth function $G:[0,1]\to \mathbb R$ we denote
\begin{equation*}
\langle \pi_{tn^{\gamma}}^n ,  G \rangle = \int_0^1 du \ G(u) \ \pi_{tn^\gamma}^n (u) = \cfrac1n\sum_{x \in \Lambda_n} \varphi_{tn^\gamma} (x) \ G(\tfrac xn) \ .
\end{equation*}
We assume that we start from some initial condition associated to a macroscopic profile $\Phi_0: [0,1] \to \mathbb R$, i.e. 
\begin{equation}
\label{eq:assiocated-profile}
\lim_{n\to \infty} \langle \pi_{0}^n ,  G \rangle = \int_0^1 du \ G(u) \,  \Phi_0 (u) \ .
\end{equation}
We expect that at any time $t>0$, we have
\begin{equation*}
\lim_{n\to \infty} \langle \pi_{tn^{\gamma}}^n ,  G \rangle = \int_0^1 du \ G(u) \, \Phi_t (u) 
\end{equation*}
where $\{\Phi_t \; ; \; t \ge 0\}$ is the solution of a suitable PDE with initial condition $\Phi_0$. 

By using the equations of motions and Eq. \eqref{eq:defK}, and the homogeneity of $K$ (i.e. $K(z)=n^{-1-\gamma} K(z/n)$), we have that
\begin{equation}
\label{eq:Dynkin}
\begin{split}
&\langle \pi_{tn^{\gamma}}^n ,  G \rangle - \langle \pi_{0}^n ,  G \rangle \\
=&\int_0^t ds\ \left\{  \frac{1}{n^2} \sum_{x,y \in \Lambda_n} K\Big( \tfrac{y-x}{n} \Big) \alpha (\varphi_{sn^\gamma} (x), \varphi_{sn^\gamma} (y)) \ G \Big( \tfrac{x}{n} \Big) \right.\\
& \left. \quad - n^{\gamma -1}  \left[ (\lambda(\Phi_\ell) + U'(\varphi_{sn^\gamma}(1)) ) G \Big( \tfrac{1}{n} \Big)- (\lambda(\Phi_r) + U'(\varphi_{sn^\gamma} (n)) G( 1) \right] \right\}\\
&+ \mathcal M_t^n (G)
\end{split}
\end{equation}
where $\mathcal M^n (G)$ is an explicit stochastic noise (martingale) such that, by a direct computation,  satisfies
\begin{equation*}
\lim_{n \to \infty} \left\langle \left[ \mathcal M_t^n (G)\right]^2 \right\rangle =0 \ .
\end{equation*}
Here $\langle \cdot \rangle$ denotes the average with respect to the initial condition and the diverse white noises involved in the dynamics.

\medskip
Assume that $G(0)=G(1)=0$. Then, since $G$ is smooth, $G(1/n)$ and $G(1-1/n)$ are of order $1/n$ and since $\gamma<2$, the second term on the right-hand side of Eq. \eqref{eq:Dynkin} vanishes. 

\medskip 
To manage the first term on the right-hand side of Eq. \eqref{eq:Dynkin} we rely on the propagation of local equilibrium assumption \cite[Section 2.3, Section 3.1]{Spohn}:

\bigskip


\noindent{\textbf{Propagation of local equilibrium (see Figure \ref{Fig:loceq}): }}{\textit{Let $\varepsilon\ll 1$. Split $\Lambda_n=\cup_{j=1}^{K_\varepsilon} \Lambda_{2\varepsilon n} (x_{j})$ into $K_\varepsilon=(2\varepsilon)^{-1}$ consecutive disjoint boxes of size $2\varepsilon n$, centred around the points $x_j$, namely $\Lambda_{2\varepsilon n} (x_j)=\{ x\in \Lambda_n \; ; \; |x -x_j| \le \varepsilon n\}$. Then at time  $tn^\gamma$, in the double limit first in $n \to \infty$ and then in $\varepsilon \to 0$, the distribution of $\{\varphi_{tn^\gamma} (z)\; ; \; z \in \Lambda_{n}\}$ is very close to{\footnote{If $\Lambda' \subset \Lambda_n$, $\mu_\Phi \Big\vert_{\Lambda'}$ denotes the marginal of $\mu_\Phi$ defined by Eq. \eqref{eq:muphidef}  when restricted to the box $\Lambda'$.}}
\begin{equation*}
\mu_{t, \varepsilon, n}^{\rm{loc}}:=  \prod_{j=1}^{K_\varepsilon} \ \mu_{\Phi_t (x_j/n)} \Big\vert_{\Lambda_{2\varepsilon n} (x_j)}\ , 
\end{equation*}
where $\Phi_t$ is the macroscopic profile we are looking for in the hydrodynamics limit{\footnote{Of course we have to assume that this property is satisfied for $t=0$, which explains the term `propagation'.}}. We recall that the Gibbs equilibrium measure $\mu_\Phi$ has been defined by Eq. \eqref{eq:muphidef}. 
}}

\bigskip
\begin{remark}
The propagation of local equilibrium assumption is a very strong assumption well accepted in physics but whose proof is one of the main problem in the mathematical theory of hydrodynamic limits \cite{Spohn, KL99}. In the case of short-range interacting particle systems in contact with reservoirs, a general theory has been developed in \cite{ELS90} for gradient stochastic lattice gas with bounded spins, following the seminal work \cite{GPV}. In the case of systems with long-range interactions, but which do not encapsulate the models under investigation, we refer the interested reader to \cite{J08, SS18} (in these papers there are no reservoirs and the systems considered are lattice gas). For systems with long range interactions in contact with reservoirs, the fact that the hydrodynamic profile satisfies the suitable boundary conditions is a non-trivial mathematical problem considered in details in \cite{BGJ21, BCGS}.
\end{remark}

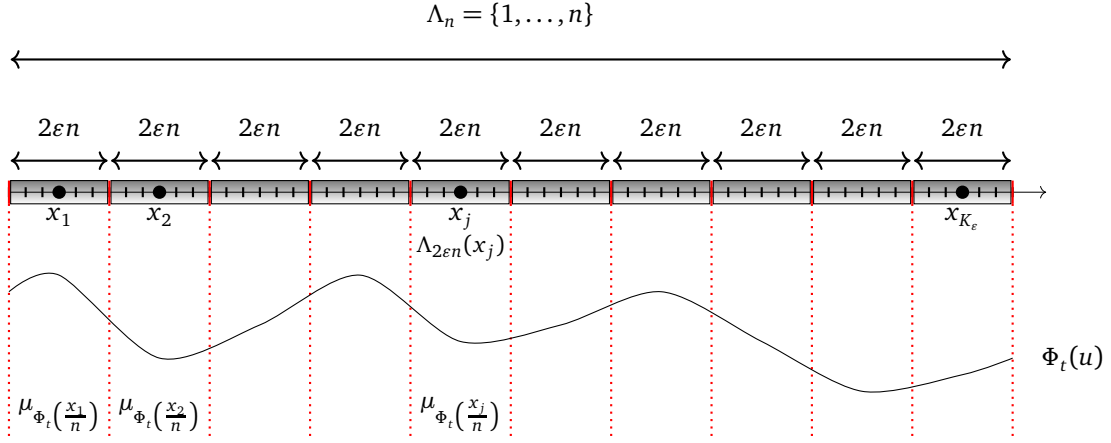
\begin{figure}
\centering
\label{Fig:loceq}
\begin{tikzpicture}[scale=0.22]

 \draw[<->,thick, black] (0.1,8) -- (59.9,8);
 \node[anchor=north] at (30,12) {$\Lambda_n=\{1,\ldots,n\}$};  

\foreach \x in {0,6,...,57} {
\shadedraw (\x+0.1,-0.7) rectangle (\x+5.9,0.7) ;
}
\foreach \x in {0,1,...,60} {
  \draw[thick] (\x,-0.3) -- (\x,0.3);}
  \draw[->,line width=0.01cm,black] (0,0) -> (62,0);
\foreach \x in {0,6,...,60} {
  \draw[thick, red] (\x,-0.7) -- (\x,0.7);
  \draw[thick,red, dotted] (\x,-0.7) -- (\x,-15);  
  }  
\foreach \x in {0,6,...,57} {
  \draw[<->,thick, black] (\x+0.1,1.9) -- (\x+5.9,1.9);
 \node[anchor=north] at (\x+3,5) {$2\varepsilon n$};}    
\coordinate (centre1) at (3,0) ;
\fill (centre1) circle (0.4) ;
\node[anchor=north] at (3,-0.4) {$x_1$};
\coordinate (centre2) at (9,0) ;
\fill (centre2) circle (0.4) ;
\node[anchor=north] at (9,-0.4) {$x_2$};
\coordinate (centre5) at (27,0) ;
\fill (centre5) circle (0.4) ;
\node[anchor=north] at (27,-0.4) {$x_j$};
\node[anchor=north] at (27,-2) {\small{$\Lambda_{2\varepsilon n} (x_j)$}};
\coordinate (centrelast) at (57,0) ;
\fill (centrelast) circle (0.4) ;
\node[anchor=north] at (57,-0.4) {$x_{K_{\varepsilon}}$};
\draw plot[smooth] coordinates {
    (0,-6) (3,-5) (9,-10) (15,-8) (21,-5) (27,-9) (33,-8) (39,-6) (45,-9) (51,-12) (57,-11) (60,-10)};
\node[anchor=east] at (66,-10) {$\Phi_t (u) $};    
\node[anchor=south] at (3,-15) {\small{$\mu_{\Phi_t \big(\tfrac{x_1}{n}\big)}$}};
\node[anchor=south] at (9,-15) {\small{$\mu_{\Phi_t \big(\tfrac{x_2}{n}\big)}$}};
\node[anchor=south] at (27,-15) {\small{$\mu_{\Phi_t \big(\tfrac{x_j}{n}\big)}$}};
\end{tikzpicture}
\bigskip

\caption{Propagation of local equilibrium. On top is represented the microscopic space $\Lambda_n$ while in the middle is represented the macroscopic space $[0,1]$ on which is living the macroscopic hydrodynamic profile $\Phi_t (u)$. On top appears the decomposition of the box $\Lambda_n$ into disjoint successive boxes $\Lambda_{2\varepsilon n} (x_j)$ of length $2\varepsilon n$ and centred around $x_j$. These boxes are microscopically large in the thermodynamic limit $n\to \infty$ but macroscopically small since $(2\varepsilon n)/n=2\varepsilon \to 0$. The Gibbs local equilibrium assumption appears at the bottom of the picture: the distribution of $\{\varphi_{tn^\gamma} (z) \; ; \; z \in \Lambda_{2\varepsilon n} (x_j) \}$ is given by the marginal (w.r.t. the box involved) of $\mu_{\Phi_t (x_j /n)}$.} 
\end{figure}

\bigskip
\begin{remark}
\label{rem:gamma2}
If $\gamma>2$, the correct time scale to observe a non-trivial time evolution of the macroscopic volume profile is the diffusive one. This has been established in \cite{BGJ21} in the context of exclusion process with long jumps. In \cite{BCGS}, a complete picture is provided for almost all complete values of $\gamma$ (the cases $\gamma \in \{1,2\}$ being pathological, they are not considered). However, extending these results to the GL dynamics considered here would require a very important work (for a generic case this would ask to develop non-gradient methods with long-range interactions in the presence of thermal baths).
\end{remark}

\bigskip
For $x\ne y$ denote the local function $\alpha_{x,y} (\varphi) = \alpha (\varphi (x), \varphi (y))$ and let 
\begin{equation}
A(\Phi,\Phi') = \int d\nu_\Phi (\phi)d\nu_{\Phi'} (\phi')  \ \alpha (\phi,\phi') 
\label{eq:A}
\end{equation}
be the expectation of $\alpha_{x,y}$ w.r.t. the product probability measure $d\nu_\Phi (\varphi(x)) \otimes d \nu_{\Phi'} (\varphi (y))$. Then, with the notations and by the propagation of local equilibrium property above, we claim then that the term appearing in Eq. \eqref{eq:Dynkin} satisfies
\begin{equation}
\label{eq:el-drift}
\begin{split}
&\lim_{n \to \infty} \int_0^t ds\ \left\{  \frac{1}{n^2}  \sum_{x,y \in \Lambda_n} K\Big( \tfrac{y-x}{n} \Big) \alpha_{x,y}  (\varphi_{sn^\gamma}) \ G \Big( \tfrac{x}{n} \Big) \right\} \\
&=\lim_{\varepsilon \to 0}  \lim_{n \to \infty} \int_0^t ds\ \left\{  \frac{1}{n^2}  \sum_{x,y \in \Lambda_n} K\Big( \tfrac{y-x}{n} \Big)  \ \left\langle \mu^{\rm{loc}}_{s,\varepsilon, n},  \alpha_{x,y} \right \rangle  \ G \Big( \tfrac{x}{n} \Big) \right\} \\
& = \int_0^t ds\ \left\{ \int_{[0,1]^2} du dv \ K(v-u)\  A(\Phi_s (u), \Phi_s (v)) G(u)  \right\} \ .
\end{split}
\end{equation}
To justify the second equality in the previous display, we split the sum as
\begin{equation}
\label{eq:localeqjust}
\begin{split}
&\frac{1}{n^2}  \sum_{x,y \in \Lambda_n} K\Big( \tfrac{y-x}{n} \Big)  \ \left\langle \mu^{\rm{loc}}_{s,\varepsilon, n},  \alpha_{x,y} \right \rangle \ G \Big( \tfrac{x}{n} \Big)\\
&= \frac{1}{n^2}  \sum_{i,j=1}^{K_\varepsilon} \sum_{x\in \Lambda_{2\varepsilon n} (i)} \sum_{y \in \Lambda_{2\varepsilon n} (j) } K\Big( \tfrac{y-x}{n} \Big)  \ \left\langle \mu^{\rm{loc}}_{s,\varepsilon, n},  \alpha_{x,y} \right \rangle  \ G \Big( \tfrac{x}{n} \Big)\\
&= \frac{1}{n^2}  \sum_{i \ne j=1}^{K_\varepsilon} \sum_{x\in \Lambda_{2\varepsilon n} (i)} \sum_{y \in \Lambda_{2\varepsilon n} (j) } K\Big( \tfrac{y-x}{n} \Big)  \ \left\langle \mu^{\rm{loc}}_{s,\varepsilon, n},  \alpha_{x,y} \right \rangle  \ G \Big( \tfrac{x}{n} \Big)\\
&+\frac{1}{n^2}  \sum_{i=1}^{K_\varepsilon} \sum_{x\ne y\in \Lambda_{2\varepsilon n} (i)} K\Big( \tfrac{y-x}{n} \Big) \  \left\langle \mu^{\rm{loc}}_{s,\varepsilon, n},  \alpha_{x,y} \right \rangle  \ G \Big( \tfrac{x}{n} \Big) \ .
\end{split}
\end{equation}
To deal with the first sum in the last equality of Eq. \eqref{eq:localeqjust} we observe that for $i \ne j$ and $x\in \Lambda_{2\varepsilon n} (i)$, $y\in \Lambda_{2\varepsilon n} (j)$, 
\begin{equation*}
\begin{split}
& \left\langle \mu^{\rm{loc}}_{s,\varepsilon, n},  \alpha_{x,y} \right \rangle  = \left\langle  \mu_{\Phi_t (x_j/n)} \big\vert_{\Lambda_{2\varepsilon n} (x_i)} \otimes \mu_{\Phi_t (x_j/n)} \big\vert_{\Lambda_{2\varepsilon n} (x_j)} \ , \  \alpha_{x,y} (\varphi)  \right\rangle \ = \  A \left(\Phi_t (x_i/n), \Phi_t (x_j/n) \right) \ ,
\end{split}
\end{equation*}
where the last equality follows from a trivial computation, remembering that $\mu_\Phi$ is a product measure with marginals given by $\nu_{\Phi}$. We also argue that in the first sum in the last equality of Eq. \eqref{eq:localeqjust} we can replace $K\Big( \tfrac{y-x}{n} \Big) G \Big( \tfrac{x}{n} \Big)$ by $K\Big( \tfrac{x_j-x_i}{n} \Big) G \Big( \tfrac{x_i}{n} \Big)$ since the difference between the two previous terms will vanish as{\footnote{In fact, this would be trivial if $K$ was a regular function but we have to be careful here because $K$ is singular at the origin. We prefer to omit this technical problem to not increase the length of the paper.}} $\varepsilon$ will go to zero. After this last replacement we recognise a discrete Riemann sum converging, as $n$ goes to infinity and then $\varepsilon$ to $0$, to
\begin{equation*}
\int_{[0,1]^2} du dv \ K(v-u)\  A(\Phi_s (u), \Phi_s (v)) G(u) \ .
\end{equation*}
Now, for the second sum in the last equality of Eq. \eqref{eq:localeqjust}, we observe that 
\begin{equation*}
\frac{1}{n^2}  \sum_{i=1}^{K_\varepsilon} \sum_{x \ne y\in \Lambda_{2\varepsilon n} (i)} K\Big( \tfrac{y-x}{n} \Big) \ \left\langle  \mu_{\Phi_s (x_i/n)} \ , \  \alpha_{x,y} (\varphi) \right\rangle  \ G \Big( \tfrac{x}{n} \Big) = 0 \ ,
\end{equation*}
since for any fixed $\Phi$, we have, thanks to the antisyammetry of $\alpha_{x,y}=-\alpha_{y,x}$, that
\begin{equation*}
\left\langle \mu_{\Phi}\ , \  \alpha_{x,y} \right\rangle =0 \ .
\end{equation*}
This concludes the justification of the second equality in \eqref{eq:el-drift}. It follows that the hydrodynamic limit of the system is given by the solution of Eq. \eqref{eq:hlGLG2}.

%
\end{proof}

\begin{remark}
The attentive reader will notice that in Eq. \eqref{eq:def-A2}, the kernel $K$ is singular and that it is not clear that the integral makes sense (in fact it does not make sense!). The precise notion of solution to Eq. \eqref{eq:hlGLG2} has to be interpreted in a weak (distributional) sense (see Appendix \ref{app:weaksolution}). 
\end{remark}

\subsection{Hydrostatic limit and properties of the NESS}
 
Taking the large time limit in the hydrodynamic equation we have that the stationary profile $\Phi_{ss} (u)$  in the NESS is the solution to 
\begin{equation}
\label{eq:hlGLG-ss}
\begin{cases}
&{\mathcal A} (\Phi_{ss}) \ (u) =0 \ ,\\
&\Phi_{ss} (0)=\Phi_\ell, \quad \Phi_{ss} (1)= \Phi_r \ .
\end{cases}
\end{equation}

\bigskip
\begin{remark}
Here also the precise notion of solution to Eq. \eqref{eq:hlGLG-ss} has to be interpreted in a weak (distributional) sense (see Appendix \ref{app:weaksolution}). 
\end{remark}


\bigskip
\begin{remark}
It is also possible to show that in the NESS the the instantaneous local rate of $\varphi$-current at $x$ defined by Eq. \eqref{eq:Jcurrent}, scales as $n^{1-\gamma}$ (for a standard diffusive system it would scale as $n^{-1}$). Indeed, the instantaneous microscopic current at $x \in \{2, \ldots, n-2\}$ was introduced in Eq. \eqref{eq:ic} and given by 
\begin{equation}
\label{eq:ic2}
J_t (x) = -\sum_{y \le x} \sum_{z \in \Lambda_n} \left\{ K (z-y) \alpha (\varphi_t (y),\varphi_t (z)) + \sqrt{K (z-y) \beta (\varphi_t (y), \varphi_t (z))} \ \dot\zeta_t (y,z)\right\}.
\end{equation}
At the boundaries the definition of the current has to be modified to take into account the exchange with the baths. By using the local equilibrium property, we deduce that the current satisfies, for any $u\in [0,1]$, 
\begin{equation*}
\lim_{n \to \infty} n^{\gamma-1} \left\langle J_t ([nu]) \right\rangle_{ss} = \int_0^u dv\  \int_{0}^1 dw \ K(w-v) \ A \left( \Phi_{ss} (v), \Phi_{ss} (w) \right) \ .
\end{equation*}
Of course, the quantity on the right hand side is falsely dependent on $u$, since taking its derivative in $u$, we get
\begin{equation*}
 \int_{0}^1 dw \ K(w-u) \  A \left( \Phi_{ss} (u), \Phi_{ss} (w) \right)
\end{equation*}
which is equal to $0$ since the profile $\Phi_{ss}$ satisfies Eq. \eqref{eq:hlGLG-ss}. 
\end{remark}

\bigskip

\subsection{Hydrodynamic limit for a time inhomogeneous perturbed dynamics}

\label{subsec:perturbed dynamics}

Consider a macroscopic time-dependent scalar field $H_t (u)$, $u\in[0,1]$, and consider the perturbed dynamics obtained by addingto the stochastic differential equation Eq. \eqref{eq:sdeB} a drift term in the form 
\begin{equation}
d_t (x,\varphi) = - \sum_{y \in \Lambda_n} K (y-x) \beta (\varphi(x), \varphi(y))\left( H_t \big(\tfrac yn\big) -H_t \big (\tfrac xn)\big) \right) \ .
\label{eq:pert}
\end{equation}
By proceeding like in a previous section (see Appendix \ref{app:HLdrifted} for details) we deduce that the hydrodynamic equations of the perturbed system are given by
\begin{equation}
\label{eq:HLdrift}
\begin{cases}
&\partial_t \Phi_t (u) - \mathcal A (\Phi_t)\ (u) \ = \ - \mathcal B_{\Phi_t} H_t \ (u)    \  ,\\
&\Phi_t (0)=\Phi_\ell, \quad \Phi_t (1)= \Phi_r \ ,\\
&\Phi_t \vert_{t=0} = \Phi_0 \ ,
\end{cases}
\end{equation}
with the linear mobility operator $\mathcal B_{\Phi_t}$ given by
\begin{equation}
\label{eq:def-operatorB}
{\mathcal B}_{\Phi_t} H_t \ (u):= \int_0^1 dv \ K(v-u) (H_t (v) -H_t (u) ) B (\Phi_t (u), \Phi_t (v))
\end{equation}
with
\begin{equation}
\label{eq:Bsquere2}
B (\Phi, \Phi') = \int d\nu_\Phi (\phi)d\nu_{\Phi'} (\phi')  \ \beta (\phi,\phi')  > 0 \ .
\end{equation}

\bigskip
\begin{remark}
The bulk part of the Markovian generator, given previously by  Eq. \eqref{eq:Gb}, has now, for the perturbed dynamics, the form
\begin{equation}
\mathcal{G}_{b,t}^{n,d}:=\mathcal{G}_{b}^{n}-\frac{1}{2}\sum_{x\in\Lambda_{n}}\sum_{y\in\Lambda_{n}}K(y-x)\beta\left(\varphi\left(x\right),\varphi\left(y\right)\right)\left(H_{t}\left(\tfrac{y}{n}\right)-H_{t}\left(\tfrac{x}{n}\right)\right)\left(\partial_{\varphi(x)}-\partial_{\varphi(y)}\right) \ .
\label{Eq:marre}
\end{equation}
Therefore  $\left[\mathcal{G}_{b,t}^{n,d}\right]^{\dagger}\left(e^{-\mathcal{E}\left(\varphi\right)}\right)\neq0$: the Gibbs probability measure ${\tilde \mu}_0 (d\varphi) \propto  e^{-\mathcal E (\varphi)} d\varphi$ is no longer invariant in the bulk for this perturbed
dynamics.  Anyway, the drift $d_{t}(x,\varphi)$ defined in Eq. \eqref{eq:pert}  conserves the `gradient
structure' given in Eq. \eqref{eq:forceF}  of the GL stochastic differential equation, i.e. the
total drift of the perturbed stochastic differential equation takes the form
\[
F_{\varphi,t}(x,y):=\exp\left(\mathcal{E}\left(\varphi\right)+\mathcal{E}_{t}^{ext}\left(\varphi\right)\right)\left(\partial_{\varphi(x)}-\partial_{\varphi(y)}\right)\left[K(y-x)\beta\left(\varphi\left(x\right),\varphi\left(y\right)\right)\exp\left(-\mathcal{E}\left(\varphi\right)-\mathcal{E}_{t}^{ext}\left(\varphi\right)\right)\right] \ ,
\]
 with the perturbed Hamiltonian
\[
\mathcal{E}_{t}^{ext}\left(\varphi\right)=-\sum_{x\in\Lambda_{n}}\varphi(x)\ H_{t}\left(\tfrac{x}{n}\right) \ .
\]
This implies that we still have $\left[\mathcal{G}_{b,t}^{n,d}\right]^{\dagger}\left(e^{-\mathcal{E}\left(\varphi\right)-\mathcal{E}_{t}^{ext}\left(\varphi\right)}\right)\neq 0$: the Gibbs probability measure proportional to $e^{-\mathcal{E}\left(\varphi\right)-\mathcal{E}_{t}^{ext}\left(\varphi\right)}d\varphi$
is an accompanying measure  (see \cite{RNCGPJS22} for physical implication of this type of measure.)
\end{remark}

\bigskip
\begin{remark}
The precise notion of solution to Eq. \eqref{eq:HLdrift} has again to be interpreted in a weak (distributional) sense (see Appendix \ref{app:weaksolution}).
\end{remark}

\subsection{Einstein relation}
 
Observe first that by Eq. \eqref{eq:forceF}, Eq. \eqref{eq:omega} (or  Eq.\eqref{eq:alpha-gamma} in this set-up), which are kind of microscopic Einstein relations,  we have the following \textit{hydrodynamic level  Einstein relation} 
\begin{equation}
\label{eq:Einstein1}
A(\Phi, \Phi') =  \left( S' (\Phi') -S' (\Phi) \right) B (\Phi, \Phi') 
\end{equation}
where $S$ is the thermodynamic entropy defined in Eq. \eqref{eq:entropy-function}. Thanks to the relation Eq. \eqref{eq:def-A2} and Eq. \eqref{eq:def-operatorB},  this implies also a second form of  \textit{hydrodynamic level  Einstein relation}: 
\begin{equation}
\label{eq:Einstein2}
\mathcal A (\Phi) = \mathcal B_{\Phi} \Big(S'(\Phi)\Big)  \ .
\end{equation}
Hence the hydrodynamic equation given by Eq. \eqref{eq:hlGLG2} becomes
\begin{equation}
\label{eq:hlGLG23}
\begin{cases}
&\partial_t \Phi_t (u) - \mathcal B_{\Phi_t (u) } \Big(S'(\Phi_t (u))\Big) = 0 \ ,\\
&\Phi_t (0)=\Phi_\ell, \quad \Phi_t (1)= \Phi_r \ ,\\
&\Phi_t \vert_{t=0} = \Phi_0 \ .
\end{cases}
\end{equation}

\bigskip
\begin{proof}[Proof of \eqref{eq:Einstein1}]
The definition \eqref{eq:A} of $A$ and Eq. \eqref{eq:alpha-gamma} give 
\begin{equation}
\begin{split}
A\left(\Phi,\Phi'\right) & =\int d\nu_\Phi(\phi) \  d\nu_{\Phi'}(\phi') \ \left[\left(U' (\phi')-U' (\phi)\right)\beta\left(\phi,\phi'\right)+\partial_{\phi} \beta \ (\phi,\phi')-\partial_{\phi'}\beta\  (\phi,\phi')\right]\\
 &=\int d\phi\  d\phi' \ 
e^{-\left(U(\phi)+\lambda(\Phi)\phi\right)+F\left(\lambda(\Phi)\right)} \ e^{- (U(\phi')+\lambda(\Phi') \phi')+F(\lambda(\Phi'))}\\
&\quad \quad \quad \quad  \quad \times \left[\left(U' (\phi')-U' (\phi)\right)\beta (\phi,\phi')+\partial_{\phi} \beta \ (\phi,\phi')-\partial_{\phi'} \beta\  (\phi,\phi')\right]
\end{split}
\end{equation}
where in the second equality we used the  definition \eqref{eq:nu_Phi}. We have then
\begin{equation}
\begin{split}
A\left(\Phi,\Phi'\right)&=
\int d\phi d\phi'\left(\frac{d}{d\phi}+\lambda({\Phi})\right)\left[e^{-\left(U(\phi)+\lambda(\Phi)\phi\right)+F\left(\lambda(\Phi)\right)} \ e^{- (U(\phi')+\lambda(\Phi') \phi')+F(\lambda(\Phi'))}  \beta(\phi,\phi')\right]\\
&-\int d\phi d\phi'\left(\frac{d}{d\phi'}+\lambda({\Phi'})\right)\left[e^{-\left(U(\phi)+\lambda(\Phi)\phi\right)+F\left(\lambda(\Phi)\right)} \ e^{- (U(\phi')+\lambda(\Phi') \phi')+F(\lambda(\Phi'))}    \beta(\phi,\phi')\right]
\end{split}
\end{equation}
and then
\begin{equation*}
A\left(\Phi,\Phi'\right)=\left[\lambda({\Phi})-\lambda({\Phi'})\right]\int d\nu_{\Phi} (\phi) d\nu_{\Phi'}(\phi') \beta \left(\phi,\phi'\right)
\end{equation*}
By the duality relation \eqref{eq:dualityLegendre} and the definition \eqref{eq:Bsquere2}, we get finally the relation  \eqref{eq:Einstein1}.

\end{proof}

\bigskip

\section{Large deviations}

\subsection{Dynamical large deviations} 
\label{sec:dynamical LD}
 
Fixing some horizon time $T>0$ and starting from an initial configuration associated to a macroscopic profile $\Phi_0$ (in the sense of Eq. \eqref{eq:assiocated-profile}), we want now to estimate the probability, as $n$ goes to infinity, to observe an atypical profile $\Phi$, i.e.
\begin{equation}
\label{eq:ldprincdyn}
\mathbb P \left( \pi_{tn^\gamma}^n \approx \Phi_t \; ; \; 0\le t \le T \right) \approx \exp(- n I_{[0,T]} (\Phi))
\end{equation}
where $\pi^n_{tn^\gamma}(u)=\tfrac1n\sum_{x \in \Lambda_n} \varphi_{tn^\gamma} (x) \delta(u-\tfrac xn)$ is the empirical volume profile at time $t n^\gamma$ and where $I_{[0,T]} (\Phi)$ is the so-called dynamical large deviations function. 
This probability will be therefore exponentially small in $n$, with however $I_{[0,T]} (\Phi)$ vanishing if and only if $\Phi$ is solution of the hydrodynamic equation \eqref{eq:hlGLG2}.

\bigskip

We prove in this section Eq. \eqref{eq:ldprincdyn} and that 
\begin{equation}
\label{eq:ratefunctdyn}
I_{[0,T]} (\Phi) = \int_0^T dt \  \mathbb L ( \Phi_t,\partial_t \Phi_t)  \ ,
\end{equation}
where the Lagrangian $\mathbb L$ takes the form
\begin{equation}
{\mathbb L} (\Phi, p) =\dfrac{1}{4} \Vert p - \mathcal A (\Phi) \Vert_{-1, {\mathcal B_\Phi}}^2   , 
\label{eq:LdevL}
\end{equation}
with the weighted fractional Sobolev norm{\footnote{While the first equality is only formal, the second one is well defined.}} defined by
\begin{equation}
\label{eq:DLDP-form-H-1form norm}
\begin{split}
\Vert \Psi \Vert^2_{-1, \mathcal B_\Phi} &:= \left\langle \Psi \ , \  (-\mathcal B_\Phi)^{-1} \Psi \right\rangle \ = \   \sup_p \left\{ \langle \Psi, p \rangle- \tfrac14 \langle p ,(-\mathcal B_\Phi) p \rangle \right\}  \ .
\end{split}
\end{equation}

 \begin{remark}
Even if we did not give the details here, it can be shown by similar arguments that the dynamical rate function ${R}_{[0,T]} (\Phi, \mathcal J)$ corresponding to the probability to observe an atypical volume profile $\Phi$ and an atypical $\varphi$-current $\mathcal J$ (satisfying the continuity equation $\partial_t \Phi_t(u)+\partial_u \mathcal J_t (u)=0 $) is given by the quadratic form 
\begin{equation*}
\begin{split}
R_{[0,T]} (\Phi,\mathcal J) & = \frac 14 \ \int_0^T dt \left\Vert \mathcal J_t - J (\Phi_t) \right\Vert^2_{-1, \mathcal B_{\Phi_t}}  \ 
\end{split}
\end{equation*}
where $J(\Phi_t)$ is the typical current given in Eq. \eqref{eq:current-form}
\begin{equation*}
J(\Phi_t (u)) =-\int_0^u dw \ \mathcal A (\Phi_t) (w) \ .
\end{equation*}
\end{remark}
\hspace{1cm}

\begin{proof}[Proof of \eqref{eq:ratefunctdyn}]
In the hydrodynamic time scale $tn^\gamma$, the ratio between the paths measure of the perturbed process $\mathbb Q^n_H$ (introduced in the previous section)  and the paths measure of the unperturbed process $\mathbb Q^n_0$ satisfies
\begin{equation}
\label{eq:Girsanov-ratio}
\begin{split}
&\cfrac{\mathbb Q^{n}_0 (\{ \pi_t \; ; \; 0\le t\le T\})}{\mathbb Q^n_{H} (\{ \pi_t \; ; \; 0\le t\le T\})} \\
&=  \exp \left\{ -\frac{n}{2}  \left[ \langle \pi_T, H_T\rangle -  \langle \pi_0, H_0\rangle -  \int_0^T  \langle \pi_t, \partial_t H_t\rangle \ dt - \int_0^T  n^\gamma \mathcal G^n  \left( \langle \pi_t, H_t\rangle\right) \ dt \right. \right.\\
& \left. \left. \quad \quad\quad\quad\quad\quad - \cfrac{1}{4} \int_0^T n^{\gamma +1} \Gamma^n \left( \langle \pi_t, H_t\rangle, \langle \pi_t, H_t\rangle \right) \ dt \right] \right\} \ . \\
\end{split}
\end{equation}
Here $\Gamma^n$ is the bilinear `carr\'e du champ' operator associated to $\mathcal G^n$ and is given, for two test functions $f,g$, by
\begin{equation}
\begin{split}
\Gamma^n (f,g) &=  \mathcal G^n (f g) - f \mathcal G^n g - g\mathcal G^n f\\
&= \sum_{x,y \in \Lambda_n} K (y-x)\ \beta (\varphi (x), \varphi(y)) \ \left[ \left( \partial_{\varphi(y)} f - \partial_{\varphi(x)} f\right) \right] \left[ \left( \partial_{\varphi(y)} g - \partial_{\varphi(x)} g\right) \right] \\
& + 2 \ \partial_{\varphi(1)} f \  \partial_{\varphi(1)} g + 2 \ \partial_{\varphi(n-1)} f \  \partial_{\varphi(n-1)} g  \ .
\end{split}
\label{eq:Gamma}
\end{equation}
In particular we have that
\begin{equation*}
\begin{split}
&\Gamma^n \left( \langle \pi_t, H_t\rangle, \langle \pi_t, H_t\rangle \right) \\
=& \frac{1}{n^2} \sum_{x,y \in \Lambda_n} K (y-x)\  \beta (\varphi (x), \varphi(y)) \ \left[  H_t \Big( \tfrac yn \Big) -H_t \Big( \tfrac xn \Big) \right]^2 + \cfrac{2}{n^2} \ \left(H_t ( 1/n )\right)^2  + \cfrac{2}{n^2} \left( H_t (1) \right)^2  \ .
\end{split}
\end{equation*}
The atypical profile $\Phi$ being fixed we choose the scalar field $H_t(u)$ such that it satisfies the linear fractional Poissson equation with homogeneous Dirichlet boundary conditions:
\begin{equation}
\label{eq:Poisson-equationG}
\begin{split}
 &- \ {\mathcal B}_{\Phi_t} H_t \ (u)= \ \partial_t \Phi_t (u) - \mathcal A (\Phi_t ) (u)\ , \quad H_t(0)=H_t(1)=0 \ .
 \end{split}
\end{equation}
Again, Eq. \eqref{eq:Poisson-equationG} has to be interpreted in a weak  sense (i.e. multiplying everything in this equation by a smooth test function with compact support, integrating in space, and using symmetry arguments to write well defined integrals) because Eq. \eqref{eq:def-operatorB} does not make sense as a classical integral. 

Thanks to this particular choice, we observe that as $n$ goes to infinity, 
\begin{equation*}
\mathbb Q^n_{H} (\{ \pi_t  \approx \Phi_t \; ; \; 0 \le t \le T\}) \approx 1 \ .
\end{equation*}
It follows that
\begin{equation*}
\begin{split}
&\mathbb P \left( \pi_{tn^\gamma}^n \approx \Phi_t \; ; \; 0\le t \le T \right) \\
&= \mathbb Q^n_{0} (\{ \pi_t  \approx \Phi_t \; ; \; 0 \le t \le T \}) = \mathbb E^n_H \left[ {\mathbb 1}_{\{ \pi_t  \approx \Phi_t \; ; \; 0 \le t \le T\}  } \ \frac{{\mathbb Q}^{n}_0 (\{ \pi_t \; ; \; 0\le t\le T\})}{{\mathbb Q}^n_{H} (\{ \pi_t \; ; \; 0\le t\le T\})} \right] \\
&=\mathbb E^n_H \left[ \frac{{\mathbb Q}^{n}_0 (\{ \pi_t \; ; \; 0\le t\le T\})}{{\mathbb Q}^n_{H} (\{ \pi_t \; ; \; 0\le t\le T\})} \right] \ .
\end{split}
\end{equation*}
We deduce, by using that for the perturbed system, the hydrodynamic profile is $\Phi$, and the propagation of local equilibrium assumption, that as $n$ goes to infinity
\begin{equation*}
\mathbb P \left( \pi_{tn^\gamma}^n \approx \Phi_t \; ; \; 0\le t \le T \right) \approx \exp(- n I_{[0,T]} (\Phi))
\end{equation*}
where the dynamical large deviations function is 
\begin{equation*}
\begin{split}
I_{[0,T]} (\Phi) & = \frac12 \int_0^T dt \ \left\{ \langle \partial_t \Phi_t - {\mathcal A} (\Phi_t) \ , \ H_t \rangle \ + \tfrac12 \langle {\mathcal B}_{\Phi_t} H_t  , H_t \rangle \right\}\\
\end{split}
\end{equation*}
if $\Phi$ satisfies $\Phi_t (0)=\Phi_\ell$, $\Phi_t (1)=\Phi_r$ for any $t\in [0,T]$ and $+\infty$ otherwise (because the probability to observe a profile which does not satisfy the Dirichlet boundary conditions will be superexponentially small in $n$).  

\bigskip

Let us now give an equivalent expression of the dynamical large deviation functional. By Eq. \eqref{eq:Poisson-equationG}, we have that
\begin{equation}
\label{eq:inverse1}
\langle \partial_t \Phi_t - {\mathcal A} (\Phi_t) \ , \ H_t \rangle = -\langle {\mathcal B}_{\Phi_t} H_t  , H_t \rangle
\end{equation}
so that
\begin{equation}
\label{eq:expressionLDF2}
\begin{split}
I_{[0,T]} (\Phi) & = \frac 14 \ \int_0^T dt \langle - {\mathcal B}_{\Phi_t} H_t  , H_t \rangle\\
&= \frac{1}{8}  \int_{[0, 1]^2} du dv \  \ K(v-u) (H_t (v) -H_t (u) )^2\  B (\Phi_t (u), \Phi_t (v)) \ .
\end{split}
\end{equation}
The last equality shows that for any scalar field $\Phi$, $-\mathcal B_{\Phi}$ is a non-negative linear operator whose kernel is constituted of constant functions. Hence restricted on the set of functions $H$ satisfying homogeneous Dirichlet boundary conditions  $H(0)=H(1)=0$, it is invertible. It is then easy to check that (use Eq. \eqref{eq:inverse1} and Eq. \eqref{eq:DLDP-form-H-1form norm})
\begin{equation}
\label{eq:DLDP-form}
\begin{split}
I_{[0,T]} (\Phi) & = \frac 14 \ \int_0^T dt \left\Vert \partial_t \Phi_t - \mathcal A (\Phi_t) \right\Vert^2_{-1, \mathcal B_{\Phi_t}}  \ .
\end{split}
\end{equation}

\end{proof}


\bigskip

\subsection{Static large deviations in the NESS : non-equilibrium free energy or quasi-potential}
\label{sec:MFT}

The MFT claims that the non-equilibrium free energy, i.e. the large deviations function of the empirical volume density in the NESS, can be recovered as the solution (called the quasi-potential)  of a variational problem involving the dynamical large deviation function $I_{[0,T]}$ introduced above. If we believe this principle is still correct we obtain hence an indirect way to compute the non-equilibrium free energy of the NESS since we have an explicit expression for $I_{[0,T]}$. \\

Classical arguments in analytic mechanics imply that the quasi-potential $V$, defined on profiles $\Phi(u)$ such that $\Phi(0)=\Phi_\ell, \Phi(1)=\Phi_r$ by
\begin{equation}
\label{eq:vf-quasipot}
V(\Phi)= \inf_{\substack{\pi(-\infty)=\Phi_{ss} \\ \pi(0)=\Phi} }\   I_{(-\infty, 0]} (\pi) = \inf_{\substack{\pi(-\infty)=\Phi_{ss} \\ \pi(0)=\Phi} }\  \int_{-\infty}^0  \mathbb L (\pi_t, \partial_t \pi_t) \ dt \ ,
\end{equation}
solves the stationary Hamilton-Jacobi equation
\begin{equation}
\label{eq:HJ-equation}
\mathbb H \left( \Phi, \tfrac{\delta V}{\delta \Phi}\right)=0, \quad  \tfrac{\delta V}{\delta \Phi} (0)= \tfrac{\delta V}{\delta \Phi} (1) =0\ 
\end{equation}
where the Hamiltonian $\mathbb  H$ is defined by
\begin{equation}
\mathbb  H(\Phi, p)= \sup_{\xi } \left\{ \langle \xi, p\rangle - \mathbb L (\Phi, \xi)\right\} = \langle {\mathcal A} (\Phi), p \rangle + \langle p, (-\mathcal B_\Phi) p \rangle \ .
\label{eq:LdevH}
\end{equation} 
Observe that by using the Einstein relation \eqref{eq:Einstein2}, it can be rewritten as
\begin{equation}
\label{eq:HJ-equation-einstein}
\left\langle \tfrac{\delta V}{\delta \Phi}, \mathcal B_\Phi \left( -S'(\Phi) + \tfrac{\delta V}{\delta \Phi}\right) \right\rangle=0, \quad  \tfrac{\delta V}{\delta \Phi} (0)= \tfrac{\delta V}{\delta \Phi} (1) =0\ .
\end{equation}
%
%
%

\bigskip
\begin{remark}
\label{rem:4.1}
Here, we simplify considerably the mathematical difficulties associated to the problem by avoiding to specify correctly the topological setting associated to this infinite-dimensional Hamilton-Jacobi equation, see \cite{BDSGLJL11} for an interesting discussion. In fact, with a complete mathematical rigour, there is not uniqueness of the solution of the stationary Hamilton-Jacobi equation but it can be shown that the non-equilibrium free energy is bigger than any stationary Hamilton-Jacobi equation vanishing for $\Phi=\Phi_{ss}$, the volume profile in the NESS (to see it it is is sufficient to follow mutatis mutandis the proof of Lemma 6.1 in \cite{BDSGLJL06}). 
\end{remark}

\bigskip
\begin{remark}
MFT claims that $V$ is the non-equilibrium free energy of the NESS \cite[Section IV]{BDSGLJL15}. While the theory is very general, it is usually very difficult to compute $V$ explicitly, even numerically. We observe that $V$ depends only on the two macroscopic parameters $A$ and $B$ appearing respectively in Eq. \eqref{eq:A2}  and Eq. \eqref{eq:Bsquere2}, which are related by the Einstein relation \eqref{eq:Einstein1}. On the other hand $V$ does not depend  on the microscopic parameters of the model.
\end{remark}

\bigskip
\begin{remark}
In Section \ref{sec:particular-case}, we will consider special cases where we can obtain the explicit maximal solution of the Hamilton-Jacobi equation \eqref{eq:HJ-equation-einstein}.
\end{remark}

\bigskip
\begin{remark}
In the literature (Eq. (4.12) in \cite{BDSGLJL15} in a diffusive set-up or \cite{DKM23} in a long-range situation), an important role is played by the so called `\textit{MFT's deterministic
equations}', which are the canonical equations associated to the Hamiltonian $\mathbb H$. This provides a theoretical way to solve the Hamilton-Jacobi equation by characteristics.  In our set-up, theses equation associated to Eq.\eqref{eq:LdevH} are given by 
\[
\begin{cases}
&\left(\partial_{t}\Phi_{t}\right)(u)= \cfrac{\delta\mathbb{H}\left[\Phi_{t},p_{t}\right]}{\delta p_{t}}\  (u) = \mathcal{A}\left[\Phi_{t}\right](u)- 2 \mathcal{B}_{\Phi_{t}}\left[p_{t}\right](u) \ , \\
&\\
&\left(\partial_{t}p_{t}\right)(u)=-\cfrac{\delta\mathbb{H}\left[\Phi_{t},p_{t}\right]}{\delta\Phi_{t}}\  (u)= - \left\langle \mathcal{A}_{\Phi_{t}}^{\ell}\left[\delta_{u}\right],p_{t}\right\rangle - \left\langle p_{t},\left(-\mathcal{B}_{\Phi_{t},u}^{\ell}\right)\left[p_{t}\right]\right\rangle \ , 
\end{cases}
\]
 with $p_t (0)=p_t(1)=0$ and the linearised operators defined by 
\[
\begin{cases}
&\mathcal{A}_{\Phi}^{\ell}\left[h\right]:=\lim_{\varepsilon\rightarrow0}\frac{\mathcal{A}\left[\Phi+\varepsilon h\right]-\mathcal{A}\left[\Phi\right]}{\varepsilon} \ ,\\
&\\
& \mathcal{B}_{\Phi,u}^{\ell} := \lim_{\varepsilon\rightarrow0}\frac{\mathcal{B}_{\Phi+\varepsilon\delta_{u}}-\mathcal{B}_{\Phi}}{\varepsilon} \ .
\end{cases}
\]
These MFT's deterministic equations are non linear non local equations, which seem a priori impossible to solve in general.
\end{remark}

\bigskip
\begin{remark}
Like in the diffusive case \cite[Section 2.4]{BDSGLJL02}, the quasi-potential $V$ is a Lyapounov function for the hydrodynamic equations. Indeed, by using the Hamilton-Jacobi equation \eqref{eq:HJ-equation} satisfied by $V$, we see that if $\Phi_t (u)$ is solution of Eq. \eqref{eq:hlGLG2} then 
\begin{equation*}
\begin{split}
\frac{d}{dt} V(\Phi_t)&= \left\langle \tfrac{\delta V}{\delta \Phi} (\Phi_t), \partial_t \Phi_t  \right\rangle = \left\langle \tfrac{\delta V}{\delta \Phi} (\Phi_t), \mathcal A (\Phi_t)  \right\rangle\\
&= - \left\langle \tfrac{\delta V}{\delta \Phi} (\Phi_t) \ , \ \left( -{\mathcal B}_{\Phi_t} \right) \tfrac{\delta V}{\delta \Phi} (\Phi_t) \right\rangle \ \le \ 0 \ ,
\end{split}
\end{equation*}
the last inequality coming from the fact that the linear operator $-{\mathcal B}_{\Phi}$ is a non-negative operator (this follows from Eqs. \eqref{eq:def-operatorB} and \eqref{eq:Bsquere2}). 

This can be seen as an extension of the second law for non-equilibrium systems. 
\end{remark}
\bigskip

\begin{remark}
At equilibrium, i.e. if $\Phi_\ell=\Phi_r ={\bar \Phi}$, the stationary profile $\Phi_{ss}$ is constant equal to $\bar \Phi$ and the quasi-potential, maximal solution of Eq. \eqref{eq:HJ-equation}, is given by 
\begin{equation}
\begin{split}
V(\Phi) &=  \int_0^1 du \ \left\{ S(\Phi (u)) -S(\bar\Phi) - S' (\bar\Phi) (\Phi (u) -\bar\Phi )\right\} \ .
\end{split}
\end{equation}
On one hand, this formula can be  directly obtained by a contraction principle from Sanov's Theorem since we know that in this case the stationary state is given by the product probability measure $\mu_{\bar \Phi}$.

On the other hand, it can be shown that $V$ defined above solves the Hamilton-Jacobi equation \eqref{eq:HJ-equation}. 
This can be proved by observing first that
\[
\tfrac{\delta V}{\delta\Phi} (u) =S'\left(\Phi(u)\right)-S'\left(\bar \Phi \right) \ ,
\] 
satisfies the boundary conditions $\tfrac{\delta V}{\delta \Phi} (0)= \tfrac{\delta V}{\delta \Phi} (1) =0$, and moreover that 
\begin{align*}
&\left\langle \mathcal{A}\left(\Phi\right),\frac{\delta V}{\delta\Phi}\right\rangle +\left\langle \frac{\delta V}{\delta\Phi},\left(-{\mathcal B}_{\Phi}\right)\frac{\delta V}{\delta\Phi}\right\rangle \\
&=\int_{0}^{1}du\ \int_{0}^{1}dv \ K(v-u)  \  \left(S'\left(\Phi(u)\right)-S'\left(\bar \Phi\right)\right) \ A (\Phi_t (u), \Phi_t (v))\\
 & -\int_{0}^{1}du\ \int_{0}^{1}dv \ K(v-u)\left(S'\left(\Phi(u)\right)-S'\left(\bar \Phi\right)\right) \ B (\Phi_t (u), \Phi_t (v))\left(S'\left(\Phi(v)\right)-S'\left(\bar \Phi\right)-\left(S'\left(\Phi(u)\right)-S'\left(\bar \Phi\right)\right)\right)\\
 & =\int_{0}^{1}du\ \int_{0}^{1}dv \ K(v-u)  \  \left(S'\left(\Phi(u)\right)-S'\left(\bar \Phi\right)\right) \ A (\Phi_t (u), \Phi_t (v))\\
 & -\int_{0}^{1}du\ \int_{0}^{1}dv \ K(v-u)\left(S'\left(\Phi(u)\right)-S'\left(\bar \Phi\right)\right) \ B (\Phi_t (u), \Phi_t (v))\left(S'\left(\Phi(v)\right)-S'\left(\Phi(u)\right)\right)\\
 & =0
\end{align*}
where the last equality comes from the Einstein relation Eq. \eqref{eq:Einstein1}. This is not sufficient to show that $V$ is really the quasi-potential but it can be shown it is the maximal solution of the Hamilton-Jacobi equation (see Remark \ref{rem:4.1}).
\end{remark}

\bigskip
\section{Diffusive limit}
\label{sec:difflimit}

\subsection{Diffusive limit of the hydrodynamics and hydrostatics}

As already mentioned in Remark \ref{rem:gamma2}, if $\gamma>2$ (resp. $\gamma=2$), we expect that the correct time scale to observe a non-trivial evolution of the macroscopic volume profile, is the diffusive one, i.e.  $tn^2$ (resp. $tn^2 / \log n$). Then we expect the dynamics will behave in fact as if it was short-range, and the hydrodynamic limits will be given by a  diffusion equation (see \cite{GS22,BCGS} for rigorous proofs in the case of the exclusion process with long jumps and \cite{DKM23} for convincing arguments in the context of Riesz gas{\footnote{In \cite{DKM23} the parameter $s$ there corresponds to $\gamma/2$ for the current paper.}}). 

However we observe that as $\gamma \to 2^-$, the {renormalised operator $\tfrac{2-\gamma}{2}{\mathcal A}$ converges to a second order differential operator (see Appendix \ref{app:weaksolution}). The reason to choose the prefactor $\tfrac{2-\gamma}{2}$ is explained in Appendix \ref{app:C21}. This shows that the hydrodynamics \eqref{eq:hlGLG2} (resp. hydrostatics \eqref{eq:hlGLG-ss}) converge as $\gamma \to 2^-$ to the solution of the parabolic equation 
\begin{equation}
\label{eq:diffusionequation2}
\partial_t \Phi_t = \partial_u \left( D (\Phi) \  \partial_ u  \Phi  \right), \quad D(\Phi)=  S^{''} (\Phi) B (\Phi,\Phi) >0  \ ,
\end{equation}
(resp. of the elliptic equation $0=\partial_u \left( D (\Phi_{ss}) \  \partial_ u  \Phi_{ss}  \right)$)
 with Dirichlet boundary conditions provided by $\Phi_\ell$ on the left and $\Phi_r$ on the right. The coefficient $B$ is defined by Eq. \eqref{eq:Bsquere2} and the thermodynamic (convex) entropy $S$ in Eq. \eqref{eq:entropy-function} (with Eq.\eqref{eq:Conv}). 
 
Finally, If $\gamma$ is formally infinite, then $K$ is formally nearest neighbour, i.e. $K(z)= {\sc1}_{|z|=1}$. Then the hydrodynamics is also in a similar diffusive form, but with a non explicit diffusion coefficient $D_{\infty}$ defined by the Green-Kubo formula, see Eqs. (1.13)-(1.15) in \cite{Q95}{{\footnote{In \cite{Q95}, to have the same notations as us, one has to replace there $h$ by $S$, $\alpha$ by $\beta/2$.}}}. More exactly, we have that 
\begin{equation}
\label{eq:D_infty}
D_\infty(\Phi)= S'' (\Phi) \ \inf_{g} \left\langle  \beta (\varphi (1), \varphi (2)) \ \left\{  1- \left[\partial_{\varphi (1)} - \partial_{\varphi (2)} \right]\zeta_g   \right\}^2\right\rangle_{ \mu_{\Phi}} \ .
\end{equation}
Here,  $\mu_{\Phi}$ is define in Eq. \eqref{eq:muphidef}, the infimum is taken over local smooth functions $g:\mathbb R^{\mathbb Z} \to \mathbb R$, i.e. which depends only on a finite number of coordinates, and $\zeta_g$ is defined by 
\begin{equation}
\label{eq:Ephisimple-2}
\zeta_g (\varphi) = \sum_y g (\tau_y \varphi)
\end{equation}
where the shift $\tau_y$ is defined via $(\tau_y \varphi)(z) =\varphi (z+y)$. By choosing $g=0$ in the previous variational formula we deduce, by a direct computation involving Eq. \eqref{eq:muphidef}, that
\begin{equation}
\label{eq:DDinf}
D_\infty (\Phi) \le D (\Phi) \ .
\end{equation}
This inequality is consistent with the fact  that the GL dynamics is less diffusive for $\gamma>2$ than for $\gamma<2$. 

\bigskip
\begin{remark}
We conjecture that for any $\gamma>2$, the hydrodynamic limit is given by a diffusion equation with the diffusion coefficient $D_\infty$  obtained  in \cite{Q95} {\footnote{Of course this requires to renormalise correctly $K$ so that the variance of $K$ is the same as for the nearest-neighbour case. }}. We also conjecture that the inequality \eqref{eq:DDinf}  is usually strict{\footnote{If $\beta$ is constant then the inequality is an equality \cite{GPV}.}}. If these conjectures are correct, this means we would have generically a $0$-th order phase transition for the diffusion coefficient at the critical value $\gamma=2$. 
\end{remark}

\subsection{Diffusive limit of the large deviations functionals}

Recall from Eq. \eqref{eq:LdevL}  that the rate functional for the dynamical large deviations of the empirical density is given, for a profile $\Phi$ satisfying the Dirichlet boundary conditions $\Phi_t(0)=\Phi_\ell$, $\Phi_t (1)=\Phi_r$,  by  
\begin{equation}
\begin{split}
I_{[0,T]} (\Phi) & = \frac 14 \ \int_0^T dt \left\Vert \partial_t \Phi_t - \mathcal A (\Phi_t) \right\Vert^2_{-1, \mathcal B_{\Phi_t}}  \ .
\end{split}
\end{equation}
In the Hamiltonian formulation, we can also write it as 
\begin{equation*}
\begin{split}
I_{[0,T]} (\Phi) & = \sup_{p} \ \int_0^T dt \ \left\{    \left\langle p_t \ , \ \partial_t \Phi_t \right\rangle  - \mathbb H (\Phi_t, p_t)   \right\}    
\end{split}
\end{equation*}
where the supremum is taken over smooth functions $p:[0,T]\times [0,1] \to \mathbb R$ and the Hamiltonian $\mathbb H$ is defined by Eq. \eqref{eq:LdevH}.
Let $\tilde \Phi$ be the renormalised profile
\begin{equation*}
{\tilde\Phi}_t = \Phi_{\tfrac{2t}{2-\gamma}} \ .
\end{equation*}
We can then rewrite $I_{[0,T]} (\Phi)$ as
\begin{equation*}
\begin{split}
I_{[0,T]} (\Phi) ={\tilde I}_{\left[0,{\tiny{\tfrac{2T}{2-\gamma}}}\right]} (\tilde \Phi) & =  \ \sup_{p} \left\{ \ \int_0^{\tfrac{2T}{2-\gamma}} du \ \left\{    \left\langle p_u \ , \ \partial_u \tilde\Phi_u - \tfrac{2-\gamma}{2}\mathcal A (\tilde\Phi_u) \right\rangle - \left\langle p_u \ , \ - \tfrac{2-\gamma}{2}\mathcal B_{\tilde\Phi_u} \, p_u \right\rangle     \right\} \right\}   \ .
\end{split}
\end{equation*}
By using the results of Appendix \ref{app:weaksolution} we conclude that the rate functional ${\tilde I}_{\left[0,T \right]} (\tilde \Phi)$ converges{\footnote{In fact, we should use $\Gamma$-convergence to be totally correct.}} to
\begin{equation}
 \ \sup_{p} \left\{ \ \int_0^T du \ \left\{    \left\langle p_u \ , \ \partial_u \tilde \Phi_u - \partial_u \left( D (\tilde \Phi_{u}) \  \partial_ u  \tilde \Phi_{u} \right)  \right\rangle - \left\langle p_u \ , \ \frac{D(\tilde \Phi_{u})}{S'' (\tilde \Phi_{u})}  \, p_u\right\rangle     \right\} \right\}  
\end{equation}
where $D$ is defined by Eq. \eqref{eq:diffusionequation2}. This coincides with the rate functional of a GL diffusive dynamics having diffusion coefficient $D$ and entropy function $S$ \cite{BDSGLJL15}. Since $D$ usually does not coincide with $D_\infty$, it is not the rate functional associated to the GL dynamics with nearest neighbour interactions. 

\bigskip
\begin{remark}
Using the definition \eqref{eq:vf-quasipot} of the quasi potential, we also deduce that the quasi-potential converges as $\gamma \to 2^-$ to the quasi-potential of a short-range GL dynamics.
\end{remark}

\bigskip
\section{Particular case}
\label{sec:particular-case}

\subsection{Additive noise}
 
In this section we consider the GL dynamics  \eqref{eq:sdeB} and \eqref{eq:Bord} where $\beta$ is constant, say $\beta=1$ to simplify,  so that by Eq. \eqref{eq:alpha-gamma}, we have
\begin{equation*}
\alpha (\varphi(x), \varphi(y))= U' (\varphi(y))- U'(\varphi(x)) \ .
\end{equation*}
Moreover, thanks to the form  $\nu_\Phi$ defined by Eq. \eqref{eq:nu_Phi}, we have   $\langle U'(\phi) \rangle_{\nu_\Phi}= S' (\Phi)$, and then Eqs. \eqref{eq:A2}, \eqref{eq:Bsquere2} become
\[
A\left(\Phi,\Phi'\right)=S'\left(\Phi'\right)-S'\left(\Phi\right) \ , \quad B\left(\Phi,\Phi'\right)=1 \ .
\]
This imply that the operators appearing in Eqs. \eqref{eq:def-A2}, \eqref{eq:def-operatorB} take the particular form 
\begin{equation*}
\label{eq:A-B-part}
\begin{split}
&\mathcal A(\Phi) (u) = \int_0^1 dv\ K(v-u)\left[ S'(\Phi(v)) - S' (\Phi(u)) \right] \ , \\
&(\mathcal B_\Phi H)\  (u) = \int_0^1 dv \ K(v-u) \left[H(v)- H(u)\right] \ .
\end{split}
\end{equation*}
It is then not difficult to see that 
\begin{equation}
\begin{split}
V(\Phi) &=\int_0^1 du \ \left( \int_{\Phi_{ss} (u)}^{\Phi (u)} \left[ S'(v) -S'(\Phi_{ss} (u))\right] \ dv \right)  \\
&=  \int_0^1 du \ \left\{ S(\Phi (u)) -S(\Phi_{ss} (u)) - S' (\Phi_{ss} (u)) (\Phi (u) -\Phi_{ss} (u) )\right\} \, 
\end{split}
\label{eq:QPE}
\end{equation}
solves the Hamilton-Jacobie equation \eqref{eq:HJ-equation}. 
This can be proved by observing first that
\[
\frac{\delta V}{\delta\Phi} (u) =S'\left(\Phi(u)\right)-S'\left(\Phi_{ss}(u)\right) \ ,
\]
and then that
\begin{align*}
&\left\langle \mathcal{A}\left(\Phi\right),\frac{\delta V}{\delta\Phi}\right\rangle +\left\langle \frac{\delta V}{\delta\Phi},\left(-{\mathcal B}_{\Phi}\right)\frac{\delta V}{\delta\Phi}\right\rangle \\
&=\int_{0}^{1}du\ \int_{0}^{1}dv \ K(v-u)\left(S'\left(\Phi(v)\right)-S'\left(\Phi(u)\right)\right)\left(S'\left(\Phi(u)\right)-S'\left(\Phi_{ss}(u)\right)\right)\\
 & -\int_{0}^{1}du\ \int_{0}^{1}dv \ K(v-u)\left(S'\left(\Phi(u)\right)-S'\left(\Phi_{ss}(u)\right)\right)\left(S'\left(\Phi(v)\right)-S'\left(\Phi_{ss}(v)\right)-\left(S'\left(\Phi(u)\right)-S'\left(\Phi_{ss}(u)\right)\right)\right)\\
 & =\int_{0}^{1}du \ \left(S'\left(\Phi(u)\right)-S'\left(\Phi_{ss}(u)\right)\right)\int_{0}^{1}dv \ K(v-u)\left(S'\left(\Phi_{ss}(v)\right)-S'\left(\Phi_{ss}(u)\right)\right)\\
 & =0
\end{align*}
where the last equality come the fact that the non-equilibrium stationary profile $\Phi_{ss}$ satisfies
\begin{equation}
0=\mathcal A(\Phi_{ss}) (u) = \int_0^1 dv\ K(v-u)\left[ S' (\Phi_{ss}(v)) - S' (\Phi_{ss}(u)) \right], \quad \Phi_{ss}(0) =\Phi_\ell, \quad \Phi_{ss} (1)=\Phi_r \ .
\label{eq:typ}
\end{equation} 

\bigskip
\begin{remark} This quasi-potential $V$ is then a local functional so that the NESS should not have long-range correlations. We expect that if $\beta$ is not constant, then the quasi-potential is non-local and that long-range correlations are present.  We will study more precisely this point in the forthcoming paper \cite{BCK24}.
\end{remark}

\bigskip
\begin{remark}
In the particular $U(\phi)=\phi^2$, the NESS is product and the non-equilibrium free energy can be computed directly \cite{BGJ24}. This can be proved similarly as in \cite{BGJS22}  where is considered the NESS of the boundary driven symmetric long-range zero-range process.  
\end{remark}
%
%
%
%
%
%

\bigskip

\subsection{Brownian energy model (BEM)}

In this section we consider a lattice conservative model which is not a GL dynamics but which becomes a GL dynamics after a simple change of variable. The BEM appeared for example in \cite{BO05,B07,B08,B08a, GKR07,GKRV09} in order to derive the Fourier's law and it has remarkable duality properties. Given the sub lattice $\Lambda_n=\{ 1, \ldots, n\}$ of $\mathbb Z$, the Brownian energy model is the Markov process $\omega_{t}:=\left\{ \omega_{t}(x)\in\mathbb{R} \; ; \; x\in\Lambda_n\right\}$ with state space $\mathbb R^{{\Lambda}_n}$ whose generator is given by 
\begin{equation}
\mathbb{G}_b^n=\frac{1}{2}\sum_{x, y\in\Lambda_n} K(y-x)\left(\omega(y)\partial_{\omega(x)}-\omega(x)\partial_{\omega(y)}\right)^{2},\label{eq:Mgen}
\end{equation}
where the kernel $K: \mathbb Z \to (0,\infty)$ is chosen to be symmetric, i.e. $ K(z)=K(-z)$.

On one hand, it is straightforward to check that for any temperature $T>0$, the Gibbs measure $\exp\left[-\frac{1}{2T}\sum_{x\in\Lambda_n}\left(\omega(x)\right)^{2}\right] \ d\omega$
is invariant (and in fact reversible) for the BEM, i.e. 
\begin{equation}
\left[\mathbb{G}_b^n \right]^{\dagger}\left(\exp\left[-\frac{1}{2T}\sum_{x\in\Lambda_{n}}\left(\omega(x)\right)^{2}\right]\right)=0 \ .
\label{eq:IMBEM}
\end{equation}
On the other hand, the BEM is not a $\omega$-conservation law as defined in the sense of Eq. \eqref{eq:conslaw1} because $\mathbb{G}_b^n \left[\omega(z)\right]=-\sum_{x\in\Lambda_n}K(x,z)\omega(z)$
and then 
$$\mathbb{G}_b^n\left[\sum_{z\in\Lambda}\omega(z)\right]=-\sum_{x\in\Lambda,z\in\Lambda}K(x,z)\omega(z) $$ 
is not identically zero, i.e. for any configuration $\omega$. Hence $\frac{d}{dt}\mathbb{E}\left[\sum_{z\in\Lambda}\omega_{t}(z)\right]$ is not identically zero for any initial configuration.

\subsubsection{BEM-GL transformation} The Brownian Energy model can be transformed into a GL model. Indeed, the energy field $\left\{ \varphi_{t}(x):=[\omega_{t}(x)]^{2}\in\mathbb{R}^{+} \; ; \; x \in \Lambda\right\}$
is also a Markovian process (thanks to Dynkin's criterium) but now with generator (see Appendix \ref{app:BEM})
\begin{equation}
{\mathcal G}_b^n=\sum_{x, y \in\Lambda_n}\left\{ K(y-x)\left(\varphi(y)-\varphi(x)\right)\left(\partial_{\varphi(x)}-\partial_{\varphi(y)}\right)+2 K(y-x)\varphi(y)\varphi(x)\left(\partial_{\varphi(x)}-\partial_{\varphi(y)}\right)^{2}\right\}\ .
\label{sympa2}
\end{equation}
This dynamics is then a peculiar case of general GL dynamics
Eq. \eqref{eq:currentintro} with $\mathcal M=\mathbb R^+$ and
\begin{equation}
\begin{cases}
F_{\varphi}(x,y)=2 K(y-x)\left(\varphi(y)-\varphi(x)\right) \ , \\
\varGamma_{\varphi}\left(x,y\right)=4 K(y-x)\varphi(y)\varphi(x) \ .
\end{cases}
\label{eq:beau2}
\end{equation}
More precisely, it is a particular case of Eq. \eqref{eq:forceF} by choosing the GL Hamiltonian $\mathcal{E}\left(\varphi\right)$ and the mobility field $\Omega_\varphi$ as
\begin{equation}
\text{\text{\text{\ensuremath{\begin{cases}
 \mathcal{E}\left(\varphi\right)=\frac{1}{2}\sum_{x\in\Lambda_n}\ln\left(\varphi(x)\right)\ ,\\
 \varOmega_{\varphi}\left(x,y\right)=\varGamma_{\varphi}\left(x,y\right)=4 K(y-x)\varphi(y)\varphi(x) \ .
\end{cases}}}}}
\label{eq:magnifique2}
\end{equation}
This follows from
\begin{align*}
&-\varOmega_{\varphi}\left(x,y\right)\left(\partial_{\varphi(x)}-\partial_{\varphi(y)}\right)\mathcal{E}\left(\varphi\right)+\left(\partial_{\varphi(x)}-\partial_{\varphi(y)}\right)\left[\varOmega_{\varphi}\left(x,y\right)\right] \\
& =-4 K(y-x)\varphi(y)\varphi(x)\ \frac{1}{2}\left(\frac{1}{\varphi(x)}-\frac{1}{\varphi(y)}\right) +4 K(y-x)\left(\varphi(y)-\varphi(x)\right) \\
 & =2 K(y-x)\left(\varphi(y)-\varphi(x)\right) \ .
\end{align*}
Therefore it corresponds then to take in Eq. \eqref{eq:alpha-gamma} and Eq. \eqref{eq:E}
\begin{equation*}
\beta(\phi,\phi')=4 \phi \phi', \quad U(\phi)=\tfrac{1}{2} \log \phi \ .
\end{equation*}
       
\bigskip
\begin{remark}
It is interesting to observe that this very particular GL dynamics also appears as an effective limit of an harmonic chain perturbed by an energy conserving noise in the weak coupling limit \cite{LO}.
\end{remark}

\bigskip
\begin{remark}
It is also possible to add to the dynamics defined by Eq. \eqref{eq:Mgen}
a transversal force (similarly to Eq. \eqref{eq:TF} for GL dynamics): 
\[
\mathbb{G}_{b}^{n}=\frac{1}{2}\sum_{x\in\Lambda_{n}}\sum_{y\in\Lambda_{n}}\left\{ K(y-x)\left(\omega(y)\partial_{\omega(x)}-\omega(x)\partial_{\omega(y)}\right)^{2}+a(y-x)\omega(x)\omega(y)\left(\omega(y)\partial_{\omega(x)}-\omega(x)\partial_{\omega(y)}\right)\right\} \ .
\]
Here $a:\mathbb Z \to \mathbb R$ is an arbitrary function{\footnote{This model has been studied in the nearest neighbour case \cite{B08a}.}}.  It is then a direct calculation to see that the Gibbs probability measure proportional to $\exp\left[-\frac{1}{2T}\sum_{x\in\Lambda_{n}}\left(\omega(x)\right)^{2}\right]$
is still invariant (like in Eq. \eqref{eq:IMBEM}) but no longer reversible. The energy field $\left\{ \varphi_{t}(x):=[\omega_{t}(x)]^{2}\in\mathbb{R}^{+} \; ; \; x \in \Lambda\right\}$ is still a peculiar choice of the general GL dynamics defined through Eq. \eqref{eq:forceF} where in the first line of Eq. \eqref{eq:beau2} we have to add the transverse drift 
\[
F_{\varphi}^{\bot}(x,y)=\left(a(y-x)-a(x-y)\right)\varphi(x)\varphi(y)
\]
which satisfies Eq. \eqref{eq:TF}. The thermodynamics properties defined in the next section are then still the same. 
\end{remark}

\subsubsection{Thermodynamical properties}  

We observe that the free energy is given, for any $\lambda>0$, by 
\begin{equation*}
F(\lambda)= -\log Z(\lambda) = - \tfrac12 \log \pi + \tfrac12 \log \lambda \ .
\end{equation*}
This follows from the simple computation
\begin{equation*}
\begin{split}
\int_{0}^\infty dx \ e^{-\tfrac{1}{2} \log x - \lambda x}&= \int_{0}^\infty dx \  x ^{-1/2} e^{ - \lambda x}= \sqrt 2 \int_0^\infty dy \  e^{- \tfrac{\lambda y^2}{2}}=\frac{1}{\sqrt 2} \int_{-\infty}^\infty dy \  e^{- \tfrac{\lambda y^2}{2}} =\sqrt{\cfrac{\pi}{\lambda}} \ .
\end{split}
\end{equation*}
Consequently, recalling Eq. \eqref{eq:entropy-function}, we get that the entropy satisfies for any $\Phi>0$
\begin{equation*}
\begin{split}
S(\Phi)=\sup_{\lambda} \left\{   F(\lambda) - \lambda \Phi \right\}= -\frac{1+\log 2 + \log \pi}{2} \ - \ \frac12 \log \Phi\ , \quad \lambda (\Phi) = \frac{1}{2\Phi} \ .
\end{split}
\end{equation*}
We have also that
\begin{equation}
\label{eq:sigmaphi234}
\sigma(\Phi) = \left\langle \phi^2 \right\rangle_{\nu_\Phi} - (\left\langle \phi \right \rangle_{\nu_\Phi})^2 = \cfrac{1}{S'' (\Phi)} =2 \Phi^2 \ .
\end{equation}

\subsubsection{Nearest neighbour case}

In the nearest neighbour case $K(z)={\sc 1}_{|z|=1}$ one easily check that, for $x \in \Lambda_n=\{1,\ldots,n\}$,
\begin{equation*}
{\mathcal G}_b^n \left[  \varphi (x) \right] = 2 \left[ \varphi(x-1) + \varphi (x+1) -2 \varphi (x) \right] \ .
\end{equation*}
Hence, in the diffusive time scale $tn^2$, the hydrodynamic equations are given by
\begin{equation}
\label{eq:HLBEM}
\begin{cases}
&\partial_t \Phi_t = 2 \partial_u^2 \Phi_t      \  ,\\
&\Phi_t (0)=\Phi_\ell, \quad \Phi_t (1)= \Phi_r \ ,\\
&\Phi_t \vert_{t=0} = \Phi_0 \ .
\end{cases}
\end{equation}
Hence, by Eqs. \eqref{eq:sigmaphi234} and \eqref{eq:HLBEM}, with the notations of \cite[Section 4]{BDSGLJL06}, we have that thermodynamical quantities are given by
\begin{equation*}
D(\Phi)=2, \quad \sigma (\Phi)=2 \Phi^2, \quad \chi(\Phi)= 2 \sigma (\Phi) =4 \Phi^2 \ .
\end{equation*}
The coefficients above are, up to irrelevant constants, the same as for the Kipnis-Marchioro-Presutti (KMP) model. In \cite{BGL15}, the authors were able to compute explicitly the quasi-potential. The same computations can therefore be reproduced for the GL dynamics derived from the BEM.   

\bigskip

\begin{remark}
In the long-range case, we have not been able to reproduce the computations of \cite{BGL15}. The reason is that the method of \cite{BGL15} relies crucially on differential calculus.
\end{remark}

\bigskip

\section{Comparison with the standard diffusive case}
\label{sec:comparison}

\subsection{GL dynamics with nearest neighbour interactions}

In order to illustrate the differences and similarities with the diffusive case, consider a GL dynamics with nearest neighbour interactions, i.e. take $\Lambda_n=\{1,\ldots, n\}$ and  Eqs.  \eqref{eq:alpha-gamma}, \eqref{eq:Bord} with  \eqref{eq:conslaw1}.

\bigskip
The hydrodynamics for the empirical volume profile \eqref{eq:ev} is given by the continuity equation{\footnote{Strictly speaking these references prove this hydrodynamic equations only with free boundary conditions.}}  \cite{ V91, Q95} (compare with Eq. \eqref{eq:hlGLG2})
\begin{equation}
    \begin{cases}
        \partial_t \Phi_t (u) + \partial_u \left[ J (\Phi_t (u))\right]=0 \ ,\\
        \Phi_t(0)=\Phi_\ell\ , \quad \Phi_t(1)=\Phi_r \ ,\\
        \Phi_t \vert_{t=0}\Phi_0  
    \end{cases}
\end{equation}
where the macroscopic instantaneous current associated to the density
profile $\Phi$ is given by (compare with Eq. \eqref{eq:current-form}) 
\begin{equation}
    J(\Phi)= - D_\infty (\Phi)\,  \partial_u \Phi \ .
    \label{eq:MCCD}
\end{equation}
Here $D_\infty (\Phi)$ is the diffusion coefficient defined by Eq. \eqref{eq:D_infty}. The values of $\Phi$ at the boundaries are fixed by some thermostats: $\Phi_\ell$ on the left and $\Phi_r$ on the right

\bigskip
The dynamical large deviations function takes then the form \textcolor{blue}{\cite{DV89,Q95, LY95, BDSGLJL15,BBP20}} (to compare with Eq. \eqref{eq:DLDP-form})
\begin{equation}
\label{eq:DLDF-USUAL}
\begin{split}
I_{[0,T]} (\Phi) & = \frac 14 \ \int_0^T dt \left\Vert \partial_t \Phi_t  + \partial_u \left[ J(\Phi_t (u))\right] \right\Vert^2_{-1,  {\mathcal B}_{\Phi_t}}  \ . 
\end{split}
\end{equation}
Here we introduced the weighted (diffusive) Sobolev norm (compare with Eq. \eqref{eq:DLDP-form-H-1form norm}) 
\begin{equation*}
\Vert \Psi \Vert^2_{-1, \mathcal B_\Phi} = \left\langle \Psi \ , \  (- {\mathcal B}_\Phi )^{-1} \Psi \right\rangle \ 
\end{equation*}
but now with the mobility operator (compare with Eq. \eqref{eq:def-operatorB})
\begin{equation}
\label{eq:mobility-operator-diff}
\mathcal B_\Phi = \partial_u\  \chi(\Phi) \  \partial_u \ ,
\end{equation}
and $\chi$ is the mobility function of the model. The latter is related to the diffusion coefficient by the associated \textit{hydrodynamic level  Einstein relation} \cite{Q95}
\begin{equation}
\label{eq:chiEinstein}
 \chi (\Phi) = \frac{D_{\infty} (\Phi)}{S'' (\Phi)}\ = \\inf_{g} \left\langle  \beta (\varphi (1), \varphi (2)) \ \left\{  1- \left[\partial_{\varphi (1)} - \partial_{\varphi (2)} \right]\zeta_g   \right\}^2\right\rangle_{ \mu_{\Phi}} \ ,
\end{equation}
where we use Eq.\eqref{eq:D_infty} in the second equality.

\medskip 
The previous Einstein relation can be equivalently stated with the macroscopic instantaneous current  Eq. \eqref{eq:MCCD} as
\begin{equation*}
-\partial_u \left[J(\Phi)\right]= {\mathcal B}_{\Phi} (S' (\Phi)) 
\end{equation*}
which has the advantage to be also the same as for GL dynamics with long-range interactions, see Eq. \eqref{eq:Einstein2} with Eq. \eqref{eq:current-form}, modulo the definition of the mobility operator $\mathcal B_\Phi$. 
\medskip

This discussion is resumed in the tabular below.

\bigskip
\begin{remark}
To compute the non-equilibrium free energy $V (\Phi)$, in the nearest-neighbour case, we have to solve a variational formula, like in Eq. \eqref{eq:vf-quasipot}, but involving now the Lagrangian appearing in Eq. \eqref{eq:DLDF-USUAL}. The quasi-potential is then the maximal{\footnote{See Remark \eqref{rem:4.1} for more explanations.}} solution of the Hamilton-Jacobi equation \eqref{eq:HJ-equation-einstein}, but with the mobility operator $\mathcal B_\Phi$ defined by Eq. \eqref{eq:mobility-operator-diff}.   This is usually an inextricable problem which has been solved only in very specific nearest-neighbour cases \cite[Section 4]{BDSGLJL06}: Exclusion process, Zero-range-process, (gradient{\footnote{In our context it corresponds to $\beta=1$.}}-)Ginzburg-Landau process, Kipnis-Marchioro-Presutti (KMP) process. The fact that this variational problem has been solved for the Zero-Range-process and the (gradient)-Ginzburg-Landau process is due to the fact that in this case the NESS is explicitly product. In the Exclusion process case, the `tour de force' was obtained in \cite{DLS02} by Derrida and coauthors thanks to the explicit representation of the NESS as a product of matrices \cite{DEHP93}. The solution for the KMP process was provided by analogy with the exclusion process \cite{BGL05} (even if the representation of the NESS is not in a known matrix product form like for the Exclusion process). 
\end{remark}

\bigskip 

\bigskip

\begin{table}[h!]
\caption{This table resumes the discussion of Section \ref{sec:comparison}}. 
\label{tab:comp}
\[
\begin{array}{|c|c|c|}
\hline
\multicolumn{3}{|c|}{}\\
\multicolumn{3}{|c|}{\text{\bf{Bulk generator dynamics in $\Lambda_n=\{1,\ldots,n\}$:}}}\\
\multicolumn{3}{|c|}{}\\
\multicolumn{3}{|c|}{\mathcal L_b^n = \frac{1}{2} \sum_{x,y \in \Lambda_n} K(y-x)\  (\partial_{\varphi (y)} -\partial_{\varphi (x)}) \left[  e^{- \mathcal E (\varphi)}\  \beta (\varphi(x), \varphi(y)) \ (\partial_{\varphi (y)} -\partial_{\varphi (x)} ) \ e^{\mathcal E (\varphi)} \right]}\\
 \multicolumn{3}{|c|}{}\\
\hline
&&\\
&\text{\bf{Diffusive case}}&\text{\bf{Super diffusive case}}\\
&&\\
\hline
&&\\
&\text{GL with nearest neighbour interactions}
&\text{GL with long-range interactions}\\
&&\\
&K(z)={\sc 1}_{|z|=1}
&K(z)={\sc 1}_{z \ne 0} |z|^{-(1+\gamma)}, \quad \gamma\in (1,2) \\
&&\\
\hline
\multicolumn{3}{|c|}{}\\
\multicolumn{3}{|c|}{\text{\bf{Hydrodynamic equations:}}}\\
\multicolumn{3}{|c|}{}\\
\multicolumn{3}{|c|}{\ \partial_t \Phi_t (u) + \partial_u \left[ J (\Phi_t (u))\right]=0, \quad \Phi_t(0)=\Phi_\ell\ , \quad \Phi_t(1)=\Phi_r , \quad \Phi_t \vert_{t=0}\Phi_0   }\\
\multicolumn{3}{|c|}{}\\
\hline 
&&\\
\text{Currents}
&  
J(\Phi(u))=-D_{\infty} (\Phi(u)) \, \partial_u \Phi(u) 
&
J(\Phi (u)) =-\int_0^u dv\  \int_{0}^1 dw \ K(w-v)\  A \left( \Phi (v), \Phi (w) \right)\\
&&\\
&D_\infty \ \text{defined in Eq.} \  \eqref{eq:D_infty} 
&A \ \text{defined in Eq.}\  \eqref{eq:A}   \\
&&\\
\hline
\text{Mobility operators}
&\begin{split}
&[\mathcal B_{\Phi} H]   (u) = \left[ \partial_u \, \chi (\Phi) \,  \partial_u \, H \right] (u)\\
& {\text{with \ }} \chi(\Phi)= \cfrac{D_{\infty} (\Phi)}{S'' (\Phi)}
\end{split}
&
\begin{split}
&\\
&[\mathcal B_\Phi H] (u)\\
&= \int_0^1 dv \ K(v-u) \ B(\Phi(u), \Phi(v)) [H(v)- H(u)]\\
& {\text{with }} \ B(\Phi,\Phi') =  \tfrac{A(\Phi,\Phi')}{S'(\Phi') - S'(\Phi)}
\end{split} \\
&&\\
\hline
\multicolumn{3}{|c|}{}\\
\multicolumn{3}{|c|}{\text{\bf{Einstein relation:}}}\\
\multicolumn{3}{|c|}{}\\
\multicolumn{3}{|c|}{-\partial_u J(\Phi)= {\mathcal B}_{\Phi} (S' (\Phi)) }\\
\multicolumn{3}{|c|}{}\\
\hline
\multicolumn{3}{|c|}{}\\
\multicolumn{3}{|c|}{\text{{\bf{Dynamical Large Deviations functional:}}}}\\
\multicolumn{3}{|c|}{}\\
 \multicolumn{3}{|c|}{I_{[0,T]} (\Phi) = \frac 14 \ \int_0^T dt \left\Vert \partial_t \Phi_t  + \partial_u J(\Phi_t) \right\Vert^2_{-1, {\mathcal B}_ {\Phi_t}}, \quad\quad  \Vert \Psi \Vert^2_{-1, \chi(\Phi)} = \left\langle \Psi \ , \  (- \mathcal B_{\Phi} )^{-1} \Psi \right\rangle}\\
\multicolumn{3}{|c|}{}\\
\hline
\multicolumn{3}{|c|}{}\\
\multicolumn{3}{|c|}{\text{\bf{Hamilton-Jacobi equation for the quasi-potential $V$ \eqref{eq:HJ-equation-einstein}:}}}\\
\multicolumn{3}{|c|}{}\\
\multicolumn{3}{|c|}{ \left\langle \tfrac{\delta V}{\delta \Phi}, \mathcal B_\Phi \left( -S'(\Phi) + \tfrac{\delta V}{\delta \Phi}\right) \right\rangle=0, \quad  \tfrac{\delta V}{\delta \Phi} (0)= \tfrac{\delta V}{\delta \Phi} (1) =0}\\  
\multicolumn{3}{|c|}{}\\
\hline
\end{array} 
\] 
\end{table}

\subsection{GL dynamics with a mix between long and short range interactions}

To conclude this comparison with the diffusive case, let us observe that we could also consider a GL dynamics where the kernel $K$ incorporates at the same time super-diffusive and diffusive effects. This was investigated in \cite{BL99} for interacting Brownian motions and in \cite{ GL97, GL98}) for lattice gas models.

 Since the diffusive effects appear only at the diffusive time scale $tn^2$, much larger that the sub-diffusive time scale $t n^\gamma$, we have to enhance the diffusive effects to see them appearing in the sub-diffusive time scale. We can therefore consider for example a kernel $K$ in the form 
\begin{equation}
K(z) = \ a n^{2-\gamma}  {\sc 1}_{|z|=1}
+ \frac{{\sc 1}_{z\ne 0}}{|z|^{1+\gamma}} 
\label{eq:Mix}
\end{equation}
where $a$ is a positive constant. In this case we expect that the macroscopic current will take the form 
\begin{equation}
\label{eq:current-form2}
J(\Phi (u)) = -a\, D_\infty (\Phi(u))\,  \partial_u \Phi (u) -\int_0^u dw \ \mathcal A (\Phi) (w)  \ .
\end{equation}
with the mobility operator $\mathcal B_{\Phi}$ acting on functions $H:[0,1] \to \mathbb R$ as
\begin{equation*}
\big[ {\mathcal B}_\Phi H \big] (u)= a\ \big[ \partial_u  \big(\chi (\Phi (u))  \ \partial_u H\big)\big] (u)
+ \int_0^1 dv \ K(v-u) \ B(\Phi(u), \Phi(v)) \big[ H(v) -H(u) \big] \ \end{equation*}
where $B$ is defined by Eq. \eqref{eq:Bsquere2} and $\chi$ by Eq. \eqref{eq:chiEinstein}. The dynamical large deviations function will take the form 
\begin{equation}
I_{[0,T]} (\Phi) = \frac 14 \ \int_0^T dt \left\Vert \partial_t \Phi_t  + \partial_u J(\Phi_t) \right\Vert^2_{-1, {\mathcal B}_ {\Phi_t}}, \quad  \Vert \Psi \Vert^2_{-1, \chi(\Phi)} = \left\langle \Psi \ , \  (- \mathcal B_{\Phi} )^{-1} \Psi \right\rangle
\label{eq:LdevM}
\end{equation}
and the quasi-potential will be again the maximal solution of the Hamilton-Jacobi equation \eqref{eq:HJ-equation-einstein}.

\bigskip
\begin{remark}
Thanks to the additive structure of \eqref{eq:Mix}, both for the hydrodynamic equation \eqref{eq:current-form2} and for the Lagrangian expression  \eqref{eq:LdevM} of the large deviation function, the structure is just additive between short-range and long-range terms. This is not longer true at the level of the Hamiltonian formulation \eqref{eq:LdevH} of the large deviations function ,  or  then, for  the non-equilibrium free energy \eqref{eq:vf-quasipot}.
\end{remark}

\bigskip
\bmhead{Acknowledgements}
The authors are grateful to J. Barr\'e, A. Dhar, M. Jara,  P. Gon\c{c}alves, A. Kundu and K. Mallick for useful discussions. This work was supported by the projects RETENU ANR-20-CE40-0005-01, LSD ANR-15-CE40-0020-01 of the French National Research Agency (ANR).

\bigskip

\appendix

\section{Proof of the form \eqref{eq:Gb} of the Bulk Markovian generator. }
\label{app:bulkG}

Due to the antisymmetry $\dot\zeta_{t}(x,y)=- \dot\zeta_{t}(y,x)$, we have
$$\left\langle d\zeta_{t}(x,y) \ d\zeta_{t}(z,w)\right\rangle =dt\left(\delta_{x,z}\delta_{y,w}-\delta_{x,w}\delta_{y,z}\right)$$
and then the Markov generator corresponding to the stochastic differential equation in Eq. \eqref{eq:sdeB}  is given by 
\begin{equation}
\begin{split}
&\mathcal{G}_{b}^{(n)}\\
&=
\sum_{\left(x,y\right)\in\Lambda_{n}} K (y-x)\left(\alpha\left(\varphi(x),\varphi(y)\right)\partial_{\varphi(x)}+\beta\left(\varphi(x),\varphi(y)\right)\partial^2_{\varphi(x)}-\beta\left(\varphi(x),\varphi(y)\right)\partial^2_{\varphi(x), \varphi(y)}\right)\\
&= \sum_{\left(x,y\right)\in\Lambda_{n}} K (y-x)\left(\alpha\left(\varphi(x),\varphi(y)\right)\partial_{\varphi(x)}+\beta\left(\varphi(x),\varphi(y)\right)\partial_{\varphi(x)}\left(\partial_{\varphi(x)}-\partial_{\varphi(y)}\right)\right)\\
&= \sum_{\left(x,y\right)\in\Lambda_{n}} K (y-x)\left(\alpha\left(\varphi(x),\varphi(y)\right)\partial_{\varphi(x)}+\frac{1}{2}\beta\left(\varphi(x),\varphi(y)\right)\left(\partial_{\varphi(x)}-\partial_{\varphi(y)}\right)^{2}\right)
\end{split}
\end{equation}
where in the last equality we used the symmetry of $K$ and $\beta$, i.e. $K(-z)=K(z)$, $\beta (\varphi (x) , \varphi(y))=\beta (\varphi (y) , \varphi(x))$. Finally, by substituting the relation \eqref{eq:alpha-gamma} in the previous display, we get
\begin{equation}
\begin{split}
\mathcal{G}_{b}^{(n)}&
=\sum_{\left(x,y\right)\in\Lambda_{n}} K (y-x)\left\{\left(e^{\mathcal{E}(\varphi)}\left(\partial_{\varphi(x)}-\partial_{\varphi(y)}\right)\left[e^{-\mathcal{E}(\varphi)}\beta\left(\varphi(x),\varphi(y)\right)\right]\right)\partial_{\varphi(x)} \right.\\
&\left. \qquad \quad  + \frac{1}{2}\beta\left(\varphi(x),\varphi(y)\right)\left(\partial_{\varphi(x)}-\partial_{\varphi(y)}\right)^{2}\right\}\\
&=\sum_{\left(x,y\right)\in\Lambda_{n}} K (y-x)\left\{\frac{1}{2}\left(e^{\mathcal{E}(\varphi)}\left(\partial_{\varphi(x)}-\partial_{\varphi(y)}\right)\left[e^{-\mathcal{E}(\varphi)}\beta\left(\varphi(x),\varphi(y)\right)\right]\right)\left(\partial_{\varphi(x)}-\partial_{\varphi(y)}\right) \right. \\
& \left. \qquad \quad +\frac{1}{2}\beta\left(\varphi(x),\varphi(y)\right)\left(\partial_{\varphi(x)}-\partial_{\varphi(y)}\right)^{2}\right\}
\end{split}
\end{equation}
where in the last equality, we used again the symmetry of $K$ and $\beta$. Finally, we obtain the relation \eqref{eq:Gb}.

\bigskip

\section{Conserved quantity of the GL dynamics and ergodicity}
\label{app:ergodicity}
Consider the GL dynamics with free boundary conditions on $\Lambda_n=\{1, \ldots,n\}$. It's generator is denoted by \eqref{eq:Gb} $\mathcal G_b^n$. 

\bigskip

Let us first show that the volume is the unique conserved quantity. A simple computation based on integration by parts shows that if $f: {\mathbb R}^{\Lambda_n} \to f(\varphi)$ is an arbitrary function then
\begin{equation}
\begin{split}
 \int_{{\mathbb R}^{\Lambda_n}} & \left( \mathcal G_b^n f \right) (\varphi)\ f(\varphi)\ d\varphi\\
 & = -\frac12\sum_{x,y \in \Lambda_n} K(y-x) \int_{{\mathbb R}^{\Lambda_n}} e^{-\mathcal E (\varphi)} \beta (\varphi(x), \varphi (y)) \left[ (\partial_{\varphi(y)} -\partial_{\varphi (x)}) f \right]^2 (\varphi) d\varphi \ .
 \end{split}
\end{equation}
If $f(\varphi)$ is a conserved quantity of the Ginzburg-Landau dynamics with free boundary conditions then $\mathcal G_b^n (f)=0$ so that for any $x,y\in \Lambda_n$, $\partial_{\varphi(y)}f =\partial_{\varphi (x)} f$, i.e there exists a constant $K$ such that for any $x\in \Lambda_n$, $\partial_{\varphi(x)} f =K$ . Let $C$ be an arbitrary constant, define the flat sub-manifold of ${\mathbb R}^{\Lambda_n}$ defined by  $E_C=\{ \varphi \in {\mathbb R}^{\Lambda_n} \; ; \; \textcolor{blue}{{\mathcal V} (\varphi)=} \sum_{x \in \Lambda_n} \varphi (x) = C \}$. Consider the restriction of $f$ to $E_C$ and observe that on $E_C$ we have{\footnote{Here the $d$ on the left-hand side denotes the total differential of $f$ and $d \varphi(x)$ the elements of the tangent space at $\varphi$ obtained as the differential of the coordinate function $\varphi \in E_C \to \varphi(x)$. }}
\begin{equation}
    df (\varphi) =\sum_{x\in \Lambda_n} \partial_{\varphi(x)}f \ (\varphi) \ d\varphi(x) = K \sum_{x\in \Lambda_n} d\varphi(x) =0
\end{equation}
where the last equality results from the definition of $E_C$. Hence there exists a function $g:{\mathbb R}^{\Lambda_n} \to \mathbb R$ such that for any configuration $\varphi$, $f(\varphi)=g({\mathcal V} (\varphi))$. This proves that the volume is the only conserved quantity of the dynamics with free boundary conditions.

\bigskip 

Let now $C\in \mathbb R$ and consider the GL dynamics restricted to the hyperplane $E_C$ defined above. If $f: E_C \to \mathbb R$ satisfied $\mathcal G_b^n f= 0$ then the previous argument shows that $f$ is a function of the volume $\mathcal V (\varphi)$ which is equal to $C$ on $E_C$. Hence $f$ is constant and the dynamics restricted to $E_C$ is ergodic.  

\bigskip

\section{Few properties of the hydrodynamic and hydrostatic equations}
\label{app:weaksolution}

\subsection{Weak solutions for the hydrodynamic and hydrostatic equations}

Here we explain how to interpret in a rigorous sense the formal hydrodynamic equation \eqref{eq:hlGLG2} and the hydrostatic equation \eqref{eq:hlGLG-ss}. Such interpretation can be adapted also for the hydrodynamic equations \eqref{eq:HLdrift} of the perturbed dynamics. This adaptation is left to the reader.  We recall that $B$ is defined in Eq. \eqref{eq:Bsquere2} and the relation between $A$ and $B$ is provided in Eq. \eqref{eq:Einstein1}. Moreover we assume the condition $B(\Phi,\Phi') \ge c>0$ for any $\Phi, \Phi'$.

\bigskip 
We say that a function $\Phi$ is a solution of the hydrodynamic equation \eqref{eq:hlGLG2} on $[0,T]$ if it bounded and satisfies the three following conditions for any $t \in [0,T]$:\\
\begin{enumerate}[1.]
\item We have that 
\begin{equation*}
\int_0^t ds \ \int_{[0,1]^2} du dv \ B (\Phi_{s} (u), \Phi_s (v)) \ K(v-u) \ \left[ S' (\Phi_{s}(v))-S'(\Phi_{s}(u)) \right]^2 \ <\ \infty \ .
\end{equation*}
\item For any continuous function $G(u)$ with compact support included into $(0,1)$ we have that{\footnote{Observe that in Eq. \eqref{eq:el-drift}, thanks to the antisymmetry of $\alpha$, we could have equivalently written the limit as $\frac 12 \int_0^t ds\  \int_{[0,1]^2} du dv \ K(v-u) A(\Phi_s (u), \Phi_s (v)) (G (u) - G(v))$.} }
\begin{equation*}
\begin{split}
&\int_0^1 du \ G (u) \Phi_t (u) - \int_0^1 du \ G (u) \Phi_0 (u) \\
&= - \frac 12 \int_0^t ds\ \left\{ \int_{[0,1]^2} du dv \ K(v-u) B(\Phi_s (u), \Phi_s (v)) \left[ S' (\Phi_s (v)) -S' (\Phi_s (u))  \right] \left[ G (v) - G(u) \right]  \right\} \ .
\end{split}
\end{equation*}
\item $\Phi_t (0)=\Phi_\ell, \quad \Phi_t (1)= \Phi_r$ \ .
\end{enumerate}
Using Cauchy-Schwarz inequality and the item $1$, we see that the term in item $2$ makes sense. We also observe that item $1$ and the assumption on $B$ implies that for any time $t$, $S' (\Phi_t)$ belongs to the factional Sobolev space $W^{\gamma/2,2}$ so that it is $(\gamma-1)/2$-H\"older and hence continuous. It follows that $\Phi_t$ is continuous and item $3$ makes sense. We expect that $\Phi_t$ is smooth on $(0,1)$ but not differentiable at the boundaries.

\bigskip 
We say that $\Phi_{ss}$ is a solution the hydrostatic equation \eqref{eq:hlGLG-ss} if it bounded and  satisfies the three following conditions:\\
\begin{enumerate}[1.]
\item We have that ($B$ is defined in Eq. \eqref{eq:Bsquere2} and the relation between $A$ and $B$ is provided in Eq. \eqref{eq:Einstein1})
\begin{equation*}
\int_{[0,1]^2} du dv \ B (\Phi_{ss} (u), \Phi_{ss} (v)) \ K(v-u) \ \left[ S' (\Phi_{ss}(v))-S'(\Phi_{ss}(u)) \right]^2 \ <\ \infty \ .
\end{equation*}
\item For any continuous function $G(u)$ with compact support included into $(0,1)$ we have that
\begin{equation*}
\begin{split}
\int_{[0,1]^2} du dv \ B (\Phi_{ss} (u), \Phi_{ss} (v)) \ K(v-u) \ \left[ S' (\Phi_{ss}(v))-S'(\Phi_{ss}(u)) \right]\ \left[ G(v) -G(u) \right ] =0 \ .
\end{split}
\end{equation*}
\item $\Phi_{ss} (0)=\Phi_\ell, \quad \Phi_{ss} (1)= \Phi_r$ \ .
\end{enumerate}

%

\subsection{Convergence of the hydrodynamic and hydrostatic equations as $\gamma \to 2^-$}
\label{app:C2}

\subsubsection{Choice of the normalisation}
\label{app:C21}

Let $G,H:[0,1] \to \mathbb R$ be smooth functions. We have that 
\begin{equation*}
\begin{split}
&\lim_{\gamma \to 2^-}  \cfrac{2-\gamma}{2} \ \int_{[0,1]^2} du dv \ K(v-u)  \left[ G(v)  - G(u)  \right] \ \left[ H (v)  - H (u)  \right] \\
&= \int_{[0,1]} du \   G' (u)  \ H' (u) \ .
\end{split}
\end{equation*}
A detailed proof is given in the next subsection (take $B=1$ and $S'(\Phi)=\Phi$ there). 
Roughly, we expand 
\begin{equation*}
\begin{split}
G(v)  - G(u) &= G' (u) \  (v-u) +o((v-u)^2) \ , \\
H(v)  - H(u) &= H' (u) \  (v-u) +o((v-u)^2) \ .
\end{split}
\end{equation*}
Hence, since $\gamma<2$, we  have, at first order, to control, for $u \in (0,1)$,  
\begin{equation*}
\int_0^1 du  \ G' (u)  \ H' (u)  \  \int_{0}^1 dv \  \vert v-u\ \vert^{1-\gamma} \ .
\end{equation*}
Observe that
\begin{equation*}
\begin{split}
\int_{0}^1 dv \  \vert v-u\ \vert^{1-\gamma} & = \int_0^u \ dz\  z^{1-\gamma} +\int_0^{1-u} dz \ z^{1-\gamma} = \frac{1}{2-\gamma} \ \left\{  u^{2-\gamma} +(1-u)^{2-\gamma} \right\}
\end{split}
\end{equation*}
so that
\begin{equation*}
\lim_{\gamma \to 2^-} \ \frac{2-\gamma}{2} \int_0^1 du  \ G' (u)  \ H' (u)  \  \int_{0}^1 dv \  \vert v-u\ \vert^{1-\gamma} = \int_0^1 du  \ G' (u)  \ H' (u)  \ .
\end{equation*}

\subsubsection{Convergence of some integral-differential operator to a second order differential operator}
Here we show that if $G:[0,1] \to \mathbb R$ is a smooth function with compact support included in $(0,1)$ and $\Phi:[0,1] \to \mathbb R$ is a smooth function on $(0,1)$, then 
\begin{equation*}
\begin{split}
&\lim_{\gamma \to 2^-}  \cfrac{2-\gamma}{2} \ \int_{[0,1]^2} du dv \ K(v-u)  B(\Phi (u), \Phi (v)) \left[ S' (\Phi (v)) -S' (\Phi (u))  \right] \left[ G (v) - G(u) \right] \\
&= \int_{[0,1]} du \  B(\Phi (u), \Phi (u)) \  S^{''} (\Phi (u)) \  \Phi' (u) \  G' (u) \ ,
\end{split}
\end{equation*}
which proves that, after a suitable renormalisation, the hydrodynamic (resp. hydrostatic) equation  \eqref{eq:hlGLG2} (resp.  \eqref{eq:hlGLG-ss}) converges as $\gamma \to 2^-$ to the solution of the parabolic equation 
\begin{equation*}
\partial_t \Phi_t = \partial_u \left( D (\Phi) \  \partial_ u  \Phi  \right), \quad D(\Phi)=  \  S^{''} (\Phi) B (\Phi,\Phi) >0  \ ,
\end{equation*}
(resp. of the elliptic equation $0=\partial_u \left( D (\Phi_{ss}) \  \partial_ u  \Phi_{ss}  \right)$)
 with Dirichlet boundary conditions provided by $\Phi_\ell$ on the left and $\Phi_r$ on the right. To see the previous convergence we write
 \begin{equation*}
\begin{split}
&\int_{[0,1]^2} du dv \ K(v-u)  B(\Phi (u), \Phi (v)) \left[ S' (\Phi (v)) -S' (\Phi (u))  \right] \left[ G (v) - G(u) \right] \\
&= \int_{|v-u| \le \varepsilon_\gamma} du dv \ K(v-u)  B(\Phi (u), \Phi (v)) \left[ S' (\Phi (v)) -S' (\Phi (u))  \right] \left[ G (v) - G(u) \right]  \\
&+\int_{|v-u| \ge \varepsilon_\gamma} du dv \ K(v-u)  B(\Phi (u), \Phi (v)) \left[ S' (\Phi (v)) -S' (\Phi (u))  \right] \left[ G (v) - G(u) \right] \ ,
\end{split}
\end{equation*}
where $\varepsilon_\gamma \to 0$ as $\gamma \to 2^-$ such that 
\begin{equation*}
\lim_{\gamma \to 2^-} \varepsilon_\gamma^{4-\gamma} = \lim_{\gamma \to 2^-} \varepsilon_\gamma^{3-\gamma}=0  \quad \text{but} \quad \lim_{\gamma \to 2^-} \varepsilon_\gamma^{2-\gamma} =1 \ .
\end{equation*}
For example we can take $\varepsilon_\gamma =\exp \left( \tfrac{\log (1-\sqrt{2-\gamma})}{2-\gamma}\right)$.

In the first integral we can replace the integrand $[ S' (\Phi (v)) -S' (\Phi (u))] \ [ G (v) - G(u)] $ by $\tfrac{d}{dv} [ S' (\Phi (v))] \vert_{v=u}\ G' (u) \ (v-u)^2$ and the error term resulting in the integral is (by Taylor expansion) of order $\int_{|v-u| \le \varepsilon_\gamma} du dv \ |v-u|^{3-\gamma} \lesssim \varepsilon_\gamma^{4-\gamma}$. The second integral can be bounded by a constant times $\int_{|v-u| \ge \varepsilon_\gamma} du dv \ |v-u|^{1-\gamma} =\int_{1 \ge |z| \ge\varepsilon_\gamma} dz \  |z|^{1-\gamma} \left(\int_0^1 du \ {\sc 1} _{0 <u+z <1} \right) \le 2 \tfrac{1- \varepsilon_{\gamma}^{2-\gamma}}{2-\gamma}$. We have to multiply these error terms by $2-\gamma$ and then they disappear. Hence as $\gamma \to 2^-$, we are left with the convergence of 
\begin{equation*}
(2-\gamma) \ \int_{|v-u| \le \varepsilon_\gamma } du dv \ B(\Phi (u), \Phi (v)) \  \frac{d}{du} [ S' (\Phi (u))] \ G' (u) \ K(v-u) \ (v-u)^2
\end{equation*}
which can be replaced by 
\begin{equation*}
(2- \gamma) \ \int_{|v-u| \le \varepsilon_\gamma } du dv \ B(\Phi (u), \Phi (u)) \  \frac{d}{du} [ S' (\Phi (u))] \ G' (u) \  \vert v-u\vert ^{1-\gamma}
\end{equation*}
with a cost at most of order $(2-\gamma) \varepsilon_\gamma^{3-\gamma}$. The last integral is equal to 
\begin{equation*}
\begin{split}
&\int_{0}^1 du\ B(\Phi (u), \Phi (u)) \  \frac{d}{du} [ S' (\Phi (u))] \ G' (u) \ \left( \int_{|z| \le \varepsilon_\gamma} dz \ |z|^{1-\gamma} {\sc 1}_{0 < u+z <1} \right)\\
&=\int_{0}^1 du\ B(\Phi (u), \Phi (u)) \  \frac{d}{du} [ S' (\Phi (u))] \ G' (u) \ \left( \int_{\sup(-u, - \varepsilon_\gamma)}^{\inf(1-u),\varepsilon_\gamma)} dz \ |z|^{1-\gamma} \right) \ .
\end{split}
\end{equation*}
For every $u \in (0,1)$ we have that 
\begin{equation*}
\lim_{\gamma \to 2^-} (2-\gamma) \ \int_{\sup(-u, - \varepsilon_\gamma)}^{\inf((1-u),\varepsilon_\gamma)} dz \ |z|^{1-\gamma} = \lim_{\gamma \to 2^-} (2-\gamma) \ \int_{- \varepsilon_\gamma}^{\varepsilon_\gamma} dz \ |z|^{1-\gamma} =2 \lim_{\gamma \to 2^-} \varepsilon_\gamma^{2-\gamma} =2 \ ,
\end{equation*}
the first equality resulting from the fact that for $\gamma$ sufficiently large, $-u <- \varepsilon_\gamma < \varepsilon_\gamma <1-u$.  By using the dominated convergence theorem, the proof is complete.  

\bigskip

\section{Derivation of the hydrodynamic limit for the perturbed dynamics}
\label{app:HLdrifted}
Our aim is to derive Eq. \eqref{eq:HLdrift} for the system where we add on each site $x\in {1,\ldots, n}$ the drift given by \eqref{eq:pert}
\begin{equation}
d_t (x,\varphi) = - \sum_{y \in \Lambda_n} K (y-x) \beta (\varphi(x), \varphi(y))\left( H_t \big(\tfrac yn\big) -H_t \big (\tfrac xn)\big) \right), 
\label{eq:pert2}
\end{equation}
so that the bulk Markovian generator of the dynamics is modified according to \eqref{Eq:marre}
while $\mathcal G_\ell$ and $\mathcal G_r$ remain the same \eqref{eq:Gbo}. The evolution of the empirical density tested against a test function $G$ is the same as Eq. \eqref{eq:Dynkin} except that in the integral appearing on the right hand side of this equation we have the extra term  
\begin{equation}
\begin{split}
   & n^{\gamma -1} \sum_{x \in \Lambda_n} d_s (x, \varphi_{sn^{\gamma}}) \ G \left( \tfrac xn \right)\\
   & = -\cfrac{1}{n^2} \sum_{x,y \in \Lambda_n} K \Big(\tfrac{y}{n}-\tfrac{x}{n}\Big) \beta (\varphi_{sn^\gamma}(x), \varphi_{sn^\gamma}(y))\left( H_s \big(\tfrac yn\big) -H_s \big (\tfrac xn)\big) \right) \ G \left( \tfrac xn \right) \ .
    \end{split}
\end{equation}
Arguing like for the derivation of Eq. \eqref{eq:el-drift} by using the local equilibrium assumption, we have that
\begin{equation}
\label{eq:el-b2}
\begin{split}
&\lim_{n \to \infty} \int_0^t ds\ \left\{  \frac{1}{n^2}  \sum_{x,y \in \Lambda_n} K\Big( \tfrac{y-x}{n} \Big) \beta (\varphi_{sn^\gamma}(x), \varphi_{sn^\gamma}(y))\left( H_s \big(\tfrac yn\big) -H_s \big (\tfrac xn)\big) \right) \ G \Big( \tfrac{x}{n} \Big) \right\} \\
& = \int_0^t ds\ \left\{ \int_{[0,1]^2} du dv \ K(v-u) B(\Phi_s (u), \Phi_s (v))\ \left[ H_s(v)-H_s(u) \right] G(u)  \right\} \ .
\end{split}
\end{equation}
Hence we get that the hydrodynamic limit is given by Eq. \eqref{eq:HLdrift}.

\bigskip
\section{Markovian dynamics for the energy field associated to the Brownian Energy Model}
\label{app:BEM}
Let us prove the form \eqref{sympa2} of the generator after the change of variables $\varphi=\omega^2$. This is proved by letting the Markovian generator (\ref{eq:Mgen}) acting 
on an arbitrary function in the form{ $f(\omega):=g(\omega^{2})=g(\varphi)$, then
\begin{equation}
\begin{split}
&\left(\mathbb G_b^n f\right) (\omega) =\frac{1}{2}\sum_{x,y \in \Lambda_n} K(y-x)\left(\omega(y)\partial_{\omega(x)}-\omega(x)\partial_{\omega(y)}\right)^{2}g(\omega^{2})\\
&=\sum_{x,y \in \Lambda_n} K(y-x)\left(\omega(y)\partial_{\omega(x)}-\omega(x)\partial_{\omega(y)}\right)\left(
\omega(y)\omega(x)\left(\partial_{\varphi(x)}g\right)(\omega^{2}) 
-\omega(x)\omega(y)\left(\partial_{\varphi(y)}g\right)(\omega^{2})
\right) \ .
\end{split}
\end{equation}
Observe now that\\

\begin{equation}
\begin{split}
&\left(\omega(y)\partial_{\omega(x)}-\omega(x)\partial_{\omega(y)}\right)\left(
\omega(y)\omega(x)\left(\partial_{\varphi(x)}g\right)(\omega^{2}) 
-\omega(x)\omega(y)\left(\partial_{\varphi(y)}g\right)(\omega^{2})
\right) \\
&=
\left(\omega(y)\right)^{2}\left(\partial_{\varphi(x)}g\right)(\omega^{2})+2\left(\omega(y)\right)^{2}\left(\omega(x)\right)^{2}\left(\partial^2_{\varphi(x)}g\right)(\omega^{2})\\
&-\left(\omega(y)\right)^{2}\left(\partial_{\varphi(y)}g\right)(\omega^{2})-2\left(\omega(y)\right)^{2}\left(\omega(x)\right)^{2}\left(\partial^2_{\varphi(x), \varphi(y)}g\right)(\omega^{2})\\
&-\left(\omega(x)\right)^{2}\left(\partial_{\varphi(x)}g\right)(\omega^{2})-2\left(\omega(y)\right)^{2}\left(\omega(x)\right)^{2}\left(\partial^2_{\varphi(x), \varphi(y)}g\right)(\omega^{2})\\
&+\left(\omega(x)\right)^{2}\left(\partial_{\varphi(y)}g\right)(\omega^{2})+2\left(\omega(y)\right)^{2}\left(\omega(x)\right)^{2}\left(\partial^2_{\varphi(y)}g\right)(\omega^{2})\\
&=\left[ \varphi(y)\left(\partial_{\varphi(x)}g\right)+2\varphi(y)\varphi(x)\left(\partial^2_{\varphi(x)}g\right) -\varphi(y)\left(\partial_{\varphi(y)}g\right)-2\varphi(y)\varphi(x)\left(\partial^2_{\varphi(x), \varphi(y)}g\right) \right]  (\varphi)\\
&+\left[- \varphi(x)\left(\partial_{\varphi(x)}g\right)-2\varphi(x)\varphi(y)\left(\partial^2_{\varphi(x), \varphi(y)}g\right) +\varphi(x)\left(\partial_{\varphi(y)}g\right)+ 2\varphi(x)\varphi(y)\left(\partial^2_{\varphi(y)}g\right) \right]  (\varphi)
\end{split}
\end{equation}
so that
\begin{equation}
\begin{split}
&\left(\omega(y)\partial_{\omega(x)}-\omega(x)\partial_{\omega(y)}\right)\left(
\omega(y)\omega(x)\left(\partial_{\varphi(x)}g\right)(\omega^{2}) 
-\omega(x)\omega(y)\left(\partial_{\varphi(y)}g\right)(\omega^{2})
\right)\\
&= \left[  \left(\varphi(y)-\varphi(x)\right)\left(\partial_{\varphi(x)}-\partial_{\varphi(y)}\right)+2\varphi(y)\varphi(x)\partial^2_{\varphi(x)} \right.\\
&\left. \quad \quad  -4\varphi(y)\varphi(x)\partial^2_{\varphi(x), \varphi(y)}+2\varphi(x)\varphi(y)\partial^2_{\varphi(y)}
\right] g(\varphi)\\
&=\left[\left(\varphi(y)-\varphi(x)\right)\left(\partial_{\varphi(x)}-\partial_{\varphi(y)}\right)+2\varphi(y)\varphi(x)\left(\partial_{\varphi(x)}-\partial_{\varphi(y)}\right)^{2}\right] g(\varphi)
\end{split}.
\end{equation}

\bigskip

\bibliographystyle{amsalpha}

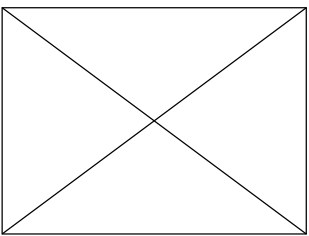
\end{document}